\documentclass[twocolumn]{aastex7}

\newcommand{\msun}{M$_\odot$}

\defcitealias{Wibking2023}{WK23}
\defcitealias{Hu2024}{H24}

\usepackage{hyperref}
\usepackage{float}
\usepackage{amsmath}
\usepackage{multirow}
\usepackage{booktabs}
\usepackage{gensymb}

\newcommand{\aref}[1]{\hyperref[#1]{Appendix~\ref{#1}}}

\begin{document}

\title{A pan-galaxy study of synthetic giant molecular filaments: a turbulence-dominated life cycle}

\correspondingauthor{Zipeng Hu, Ke Wang}

\author[orcid=0000-0002-3758-552X,sname='Hu']{Zipeng Hu}
\altaffiliation{Boya Fellow}
\affiliation{Kavli Institute for Astronomy and Astrophysics, Peking University, 5 Yiheyuan Road, Haidian District, Beijing 100871, People’s Republic of China}
\email[show]{zphu.charles@gmail.com}

\author[orcid=0000-0002-7237-3856,sname='Wang']{Ke Wang}
\affiliation{Kavli Institute for Astronomy and Astrophysics, Peking University, 5 Yiheyuan Road, Haidian District, Beijing 100871, People’s Republic of China}
\email[show]{kwang.astro@pku.edu.cn}

\author[orcid=0000-0003-3893-854X,sname='Krumholz']{Mark R. Krumholz}
\affiliation{Research School of Astronomy and Astrophysics, Australian National University, Canberra ACT 2601, Australia}
\email{mark.krumholz@anu.edu.au}

\author[orcid=0009-0003-2243-7983,sname='Su']{Keyun Su}
\affiliation{Kavli Institute for Astronomy and Astrophysics, Peking University, 5 Yiheyuan Road, Haidian District, Beijing 100871, People’s Republic of China}
\email{suky@stu.pku.edu.cn}

\begin{abstract}

Recent surveys of the Galactic plane have revealed dozens of giant molecular filaments (GMFs), with lengths ranging from tens to hundreds of parsecs, yet their origins and life cycles remain debated. In this work, we analyze over 700 GMFs identified from synthetic CO emission maps of a high-resolution magnetohydrodynamic simulation of a Milky Way–like galaxy, whose lengths range from $\sim$ 10 pc to $\sim$ 300 pc. We find that turbulent shock from galactic shear and stellar feedback are the primary drivers of GMF formation. Magnetized turbulence dominates their internal dynamics, supporting the filaments against global collapse while simultaneously inducing fragmentation into dense clumps. This fragmentation follows the turbulence-driven sausage instability model, rather than pure Jeans instability, and triggers efficient star formation along the filaments. Cloud–cloud collisions are frequent, affecting more than 70\% of GMFs, and often disrupt or reshape their morphology. The typical filamentary lifetime is $t_{\mathrm{fil}} = 14^{+2}_{-5}$ Myr, comparable to the crossing time of giant molecular clouds (GMCs). The molecular gas half-life is $\sim 7$ Myr, similar to that of GMCs, indicating that GMFs are transient but dynamically important structures.

\end{abstract}



\section{Introduction}

Over the past two decades, high-resolution surveys with ALMA, JWST, VLA, and \textit{Herschel} have firmly established that the interstellar medium (ISM) in our Galaxy and nearby systems is organized into hierarchical networks of filamentary structures \citep[e.g.,][]{Andr2010, Arzoumanian2011, Andr2014, Wang2015, Hacar2018, LiC2021_Cattail, Pare2024, Temim2024, Wang2024, JiangYu2025}. These structures range from kiloparsec-long atomic filaments to sub-parsec dense fibers within molecular clouds \citep{Hacar2022}. This filamentary web serves as the primary nursery for star formation, connecting large-scale gas reservoirs to the dense clumps and cores where stars are born. Among these structures, Giant Molecular Filaments (GMFs) represent the largest, high–aspect-ratio filaments ($>10$ pc) that are dense enough to host star formation. Since the discovery of Nessie \citep{Jackson2010, Goodman2014}, a $\sim 150$ pc filament with an aspect ratio over 300:1, dozens of GMFs have been identified across the Galactic plane \citep{Ragan2014, Wang2015, Zucker2015, Wang2016, Abreu-Vicente2016, Ge2022, GeYF2023, Wang2024}. Owing to their large spatial extent and high masses, GMFs both connect galactic dynamics to star formation and contribute significantly to the Galactic star formation rate. Understanding their life cycle and the underlying physical mechanisms is therefore essential for developing a complete theory of star formation and galactic evolution, and has become increasingly timely with the advent of high-resolution GMF catalogs.

A central but unresolved question concerns which physical processes govern the life cycle of GMFs. Early studies of Nessie and a handful of other GMFs \citep{Goodman2014, Wang2015, Zucker2015} reported strong associations with spiral arms, suggesting that GMFs form the “bones” of spiral arms and are shaped by the spiral gravitational potential throughout their evolution. However, other surveys find such associations less frequent \citep{Ragan2014, Zucker2018, Wang2024}, challenging the universality of this picture. Owing to their large sizes, GMFs are subject to a complex interplay of processes—including self-gravity, magnetic fields, turbulence, stellar feedback, and cloud–cloud collisions. The relative importance of each mechanism at different evolutionary stages remains uncertain. Statistical studies of the relative orientation between filament spines and magnetic fields have shown a tendency for alignment at low column densities, transitioning to preferentially perpendicular orientations at high column densities \citep{Planck2016, Jow2018_PRS}. This transition, reported at $N \simeq 10^{21-22}$ cm$^{-2}$, has been interpreted as evidence that gas accretion flows are guided by magnetic fields during the early formation stage \citep{Soler2017}. However, using \textit{Planck} data, \citet{Wang2024} reported a random distribution of relative orientations in a linear filament catalog across the full Galactic plane, suggesting that multiple formation mechanisms may be at play. Achieving a clear picture of the GMF life cycle requires precise theoretical modeling that incorporates realistic galactic environments and physical processes.

Recent multiscale galactic simulations have begun to address this complexity \citep{Duarte-Cabral2017, Zucker2019, Zhao2024, Pillsworth2025}, but several limitations remain. First, in those works most GMF samples are extracted from kpc-scale subregions, which may not capture the diversity of galactic environments. Second, synthetic GMFs are often identified directly from volume or column density thresholds applied to simulation data, creating a methodological gap with observations, which rely on 2D projections and specific tracers such as CO or dust emission. Finally, crucial physical mechanisms are often omitted. For example, \citet{Zucker2019} excluded self-gravity, magnetic fields, and stellar feedback, while \citet{Duarte-Cabral2017} neglected magnetic fields altogether. None of these works included pre-supernova feedback such as photoionization, which can rapidly disperse GMCs before supernova explosions occur \citep{Grudic2018, Kruijssen2019, Chevance2023}.

In this study, we overcome these limitations by analyzing a high-resolution magnetohydrodynamic (MHD) simulation of an entire Milky Way–like galaxy that incorporates a comprehensive stellar feedback model. We apply a filament identification algorithm directly to synthetic CO emission maps generated from the simulation, mirroring observational methodologies. This approach enables a robust and self-consistent investigation of the GMF life cycle within a realistic, dynamically evolving galactic environment.

This paper is organized as follows. In \autoref{sec:methods}, we describe the simulation setup, the implementation of key physical mechanisms, the synthetic CO pipeline, and the GMF identification algorithm. In \autoref{sec:resultsA}, we present the statistical properties of the simulated GMF population and compare them with observational catalogs. In \autoref{sec:resultsB}, we analyze the full GMF life cycle, quantifying the roles of different physical mechanisms across evolutionary stages. In \autoref{sec:discussion}, we discuss how different GMF identification algorithms affect the GMF properties, and the difference between GMFs and non-filamentary giant molecular clouds (GMCs). We summarize our findings in \autoref{sec:conclusion}.

\section{Methods}
\label{sec:methods}
\autoref{sec:simulation} describes the basic setup of our Milky Way–like galaxy simulation and the implementation of the key physical mechanisms in our numerical model. \autoref{sec:synthetic_obs} summarizes the pipeline for generating synthetic CO emission maps, and \autoref{sec:GMF_identify} outlines the GMF identification algorithm.

\subsection{Simulation}
\label{sec:simulation}

The numerical data analyzed in this study are drawn from the Milky Way–like galactic simulation of \citet[hereafter \citetalias{Hu2024}]{Hu2024}. The simulation was performed with the \textsc{GIZMO} code, which employs a mesh-free, Lagrangian finite-mass Godunov method to solve the magnetohydrodynamic (MHD) equations \citep{Hopkins2015, Hopkins2016a, Hopkins2016b}. Below we provide a brief overview of the setup and numerical methods, and refer readers to \citetalias{Hu2024} for full details. The initial conditions follow the “high-resolution” case of the \textsc{AGORA} project \citep{Kim2016}, which includes a dark matter halo of mass $M_\mathrm{DM} = 1.07 \times 10^{12}$ \msun, a stellar disk of mass $M_* = 3.4 \times 10^{10}$ \msun, a bulge of mass $M_\mathrm{B} = 4.3 \times 10^9$ \msun, and a gas disk of mass $M_\mathrm{gas} = 8.6 \times 10^9$ \msun. Thus, the gas mass constitutes $\sim$20\% of the total baryonic mass. The gas disk has a scale length of $R_0 = 3.43218$ kpc and a scale height of $z_0 = 0.343218$ kpc. The initial magnetic field of the simulated galaxy in cylindrical coordinates ($R, \; \phi, \; z$) can be described as
\begin{flalign}
& B_R = 0, \\
& B_\phi = B_0 \; \mathrm{exp}(-R/R_0) \; \mathrm{exp}(-|z|/z_0), \\
& B_z = 0,
\end{flalign}
where $B_0 = 10 \; \mathrm{\mu G}$. The toroidal field geometry and the mean field strength are chosen to mimic the ordered component of the observed Galactic magnetic field \citep{Beck2015}.

\begin{figure*}
\begin{center}
\includegraphics[width=\textwidth]{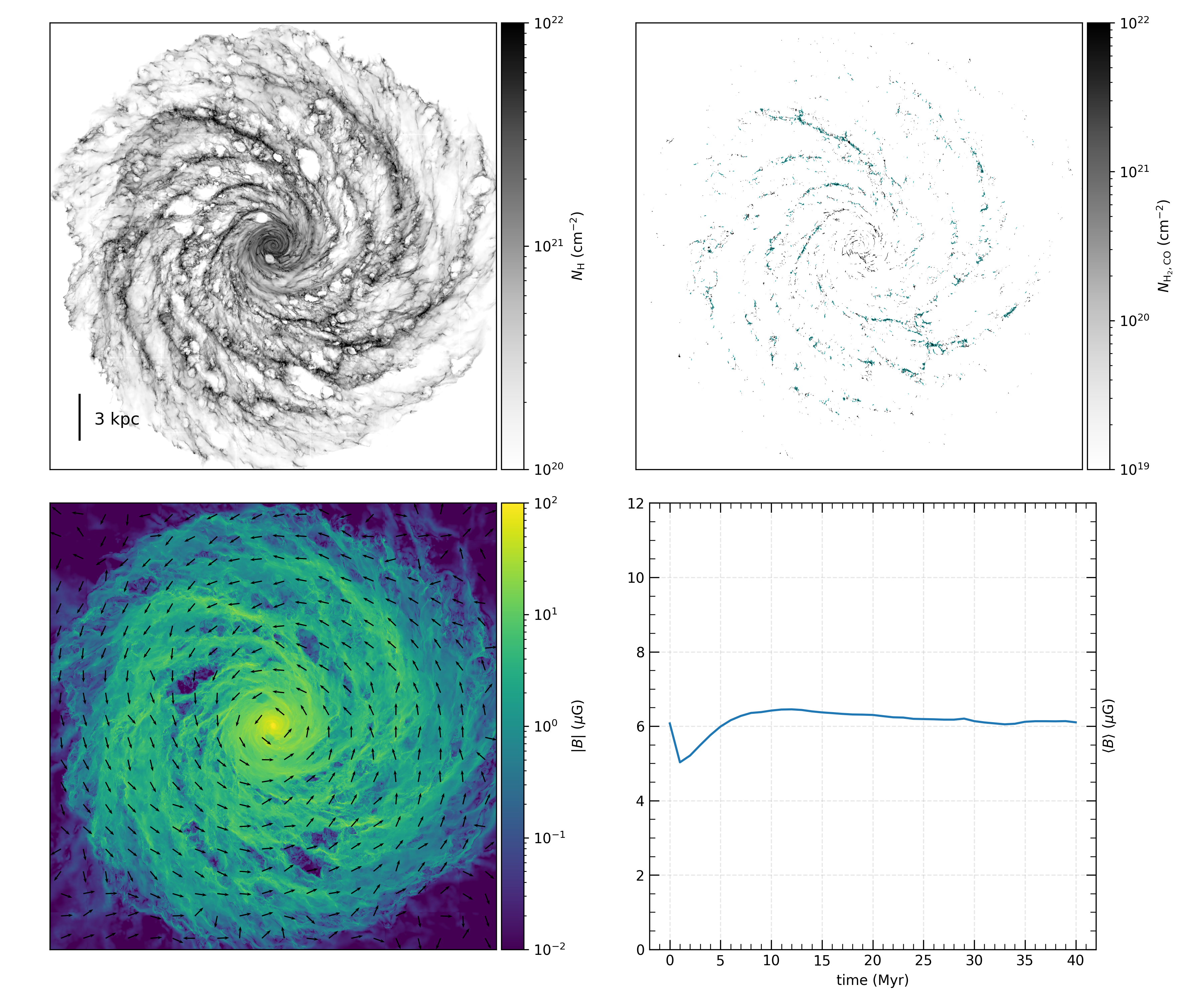}
    \caption{Milky Way–like spiral galaxy simulation at $t = 720$ Myr. Upper left: total gas column density map, where $N_\mathrm{H}$ is the total column density of H nuclei. Upper right: molecular hydrogen column density traced by synthetic CO $J = 1 \rightarrow 0$ emission. The cyan contours illustrate the positions of GMFs identified in \autoref{sec:GMF_identify}. Each map covers $30 \times 30$ kpc with a pixel size of 1 pc. Lower left: mass weighted B field strength map. The green arrows indicate the local magnetic field directions in the x-y plane. Lower right: the mass-weighted mean magnetic field as a function of time in the second stage of this simulation.}
\label{fig: 1 galactic maps}
\end{center}
\end{figure*}

The simulation consists of two stages. In the first stage, it is evolved for 0.7 Gyr at a gas mass resolution of $\Delta M_\mathrm{gas} = 859.3$ \msun, allowing the galaxy to settle into a stable star formation rate (SFR) of $\sim 3$ \msun yr$^{-1}$. At the end of this stage, all gas particles are split into ten equal-mass particles, improving the mass resolution to $\sim 90$ \msun. This corresponds to a Jeans length of $\sim 0.6$ pc, which serves as the characteristic spatial resolution. The snapshot produced after this particle splitting becomes the initial condition of the second stage, which is run for 40 Myr—longer than the 23 Myr duration in \citetalias{Hu2024}—to ensure that the full life cycle of GMFs can be traced. This extension is motivated by previous studies, which report typical GMC lifetimes of $\sim 20$ Myr \citep{Kruijssen2019, Chevance2020}. Snapshots are output every 1 Myr. For our analysis, we focus on the snapshot at $t = 720$ Myr, which provides a sufficient time window (20 Myr) both before and after for tracing GMF formation and subsequent evolution. In \autoref{fig: 1 galactic maps}, the face-on gas surface density of this snapshot is shown in the upper left panel, and the face-on magnetic field strength map is shown in lower left panel. Previous galaxy simulation works show that the field strength would be amplified by turbulence and stellar feedback in the early stage of galaxy evolution, then reaches a steady state \citep[e.g.,][]{Ntormousi2018, Robinson2025}, which would affect the star formation process and mass distribution within GMCs Therefore, we plot the mass-weighted mean magnetic field strength in the second simulation stage against simulation time in the lower right panel of \autoref{fig: 1 galactic maps}, which confirms that the galactic magnetic field in this simulation stage has reached a steady state.

The \textsc{GIZMO} code incorporates a time-dependent chemical network for hydrogen and helium cooling, together with metal-line cooling computed from pre-tabulated \textsc{CLOUDY} models \citep{Ferland1998} via the \textsc{GRACKLE} library \citep{Smith2017}. Star formation is modeled stochastically by converting gas particles into star particles once their density exceeds $\rho_{\rm crit} = 1000 \, m_{\rm H}$ cm$^{-3}$, with an efficiency per free-fall time of $\epsilon_\mathrm{ff} = 0.01$. To limit computational expense, extremely dense gas ($\rho_{\rm g} > 100 \rho_{\rm crit}$) is converted rapidly into stars by assigning $\epsilon_\mathrm{ff} = 1$. Because the stellar particle mass is only $\sim 89$ \msun, too low to fully sample the initial mass function (IMF), stellar feedback is implemented stochastically. For each star particle, the stellar population synthesis code \textsc{SLUG} \citep{Da_Silva2012, Krumholz2015} samples a unique stellar population from a Chabrier IMF \citep{Chabrier2005}.  

We focus on two kinds of stellar feedback most relevant at parsec scales: Type II supernovae (SNe) and photoionization. For SNe, \textsc{SLUG} provides the number of events ($N_\mathrm{SNe}$) and the total ejecta mass ($M_\mathrm{ej}$) at each timestep. Assuming each event releases $10^{51}$ erg, we calculate the total ejected energy and momentum, which are then deposited into neighboring gas particles following the algorithm of \citet{Hopkins2018a, Hopkins2018b}. For photoionization we use a Str\"omgren volume method, whereby the ionizing luminosity from \textsc{SLUG} is used to sequentially heat the nearest non-ionized ($T < 10^4$ K) gas particles to $10^4$ K until the photon budget is exhausted. See \citetalias{Hu2024} for full details on how SNe and photoionization are implemented.

\subsection{Synthetic observation}
\label{sec:synthetic_obs}

To mimic how observers identify dense molecular structures, we generate synthetic CO emission observations of the entire simulated galaxy. We adopt the same algorithm as in \citetalias{Hu2024}, and therefore only provide a brief summary here. First, we post-process our simulation snapshots using the astrochemistry and radiative transfer code \textsc{despotic} \citep{Krumholz2014}. This approach is computationally more efficient than performing on-the-fly chemistry calculations and has the important advantage that it captures non-LTE (Local Thermodynamic Equilibrium) excitation effects, which must be included in order to produce credible CO emission predictions. Using \textsc{despotic}, we pre-compute a look-up table of the $^{12}$CO $J=1 \to 0$ luminosity per hydrogen atom, $l_{\rm CO}$. The table spans a grid in the parameter space of three key local gas properties: hydrogen number density ($n_{\rm H} = 10^{-2} - 10^{6}~\rm cm^{-3}$), column density ($N_{\rm H} = 10^{19} - 10^{25}~\rm cm^{-2}$), and velocity gradient ($dv/dr = 10^{-3} - 10^2~\rm km~s^{-1}~pc^{-1}$). The interstellar radiation field (ISRF) strength is assumed to be solar neighborhood value throughout the computation.

For each gas particle in the simulation, we obtain $n_{\rm H}$ and $dv/dr$ directly from the simulation outputs, while $N_{\rm H}$ is derived from the corresponding pixel in the face-on column density map shown in \autoref{fig: 1 galactic maps} (upper left panel). We then determine $l_{\rm CO}$ by performing a trilinear interpolation on the pre-computed grid. The total luminosity of each particle is computed as
\begin{equation}
L_{\rm CO} = \frac{X_{\rm H} m_{\rm g} l_{\rm CO}}{m_{\rm H}},
\end{equation}
where $X_{\rm H} = 0.76$ is the total hydrogen mass fraction (including all phases of hydrogen) and $m_{\rm g}$ is the gas particle mass. Gas with an ionized fraction greater than 10\% is assigned $L_{\rm CO} = 0$, since our grid does not include the effects of photoionization.  

To enable comparison with observations, we convert this luminosity to an inferred molecular hydrogen mass using
\begin{equation}
\frac{m_{\rm H_2, CO}}{M_\odot} = 2.3 \times 10^{-29} \frac{L_{\rm CO}}{\rm erg~s^{-1}},
\end{equation}
which incorporates both the mass-to-luminosity conversion factor $X_\mathrm{CO}$ from \cite{Bolatto2013} and the $^{12}$CO $J = 1 \to 0$ line-luminosity conversion factor from \cite{Solomon2005}. Both values are fitted from solar neighborhood observations. This molecular mass can be further converted into the number of hydrogen molecules. In \autoref{fig: 1 galactic maps} (upper right panel), we present the face-on synthetic $N_{\rm H_2, CO}$ column density map for the 720 Myr snapshot. The pixel size is chosen to be 1 pc, approximately twice the characteristic spatial resolution of our simulation, and this resolution is maintained for all synthetic observations in this work.

As is common practice in observational studies \citep[e.g.,][]{Ragan2014, Abreu-Vicente2016}, we adopt a constant $X_\mathrm{CO}$ conversion factor throughout the simulated galaxy. This approach helps to minimize discrepancies when comparing our synthetic Giant Molecular Filament (GMF) catalog with observational catalogs. However, there is strong theoretical and observational evidence that $X_\mathrm{CO}$ varies systematically with gas metallicity \citep[e.g.,][]{Glover11b, Feldmann12a, Bolatto13a, Gong2020} and other physical properties such as temperature, velocity dispersion, and density structure \citep{Shetty11a, Shetty11b, Narayanan11a, Narayanan12a, Ramambason2024}.
Such environmental and structural dependencies introduce significant uncertainties into GMF mass measurements inferred from previous observational surveys. Due to current computational constraints, our simulations assume constant solar neighborhood values for the metallicity and ISRF strength rather than modeling their time-dependent evolution, an assumption that is consistent with the methodology of our synthetic observation algorithm. Consequently, the synthetic GMF catalog identified in \autoref{sec:GMF_identify} is considered reliable under these constant conditions but may not fully capture the influence of a varying galactic environment on the derived masses.

\subsection{GMF identification}
\label{sec:GMF_identify}

The most common method used in previous observational studies to identify GMFs is to apply simple column density or integrated intensity thresholds to locate dense GMCs, followed by visual inspection to identify filamentary structures \citep{Ragan2014, Wang2015, Zucker2015, Zucker2018, Abreu-Vicente2016}. Due to the large area of our simulated galaxy, we employ automated algorithms rather than visual inspection to search for filaments. Algorithms used in prior studies include \textsc{DisPerSE} \citep{Sousbie2011}, \textsc{SCIMES} \citep{Colombo2015},  \textsc{FilFinder} \citep{Koch2015}, and customized minimum spanning tree \citep[\textsc{MST}][]{Wang2016,Wang2021_MSTcode}. However, \textsc{DisPerSE} and \textsc{SCIMES} have been reported to have limited effectiveness in detecting elongated structures on large scales \citep{Zucker2019}, and \textsc{MST} requires a clump catalog as input. Therefore, we adopt \textsc{FilFinder} to identify GMFs in our synthetic observations. \textsc{FilFinder} can uniformly identify filamentary structures over a wide dynamic range of intensities and can measure filament properties such as lengths, widths, and radial profiles.

We apply \textsc{FilFinder} in two different ways: one is to a face-on view of the galaxy as shown in the upper right panel of \autoref{fig: 1 galactic maps}, and the other is to an edge-on view as in real observations. These approaches are complementary in that the former lets us identify filaments more accurately and fast by eliminating line of sight confusion\footnote{In principle we could also attempt to identify filaments in the full three-dimensional galaxy simulation using the \textsc{FilFinder3D} module, which would provide even more accurate ``true'' filaments. However, we found that applying this package to a simulation as large as ours was prohibitively computationally expensive.}, while the latter is a truer representation of the process used to produce observed GMF catalogs.

The first step for the face-on case is to apply a mask to the face-on synthetic $N_\mathrm{H_2, CO}$ column density map shown in \autoref{fig: 1 galactic maps}. Previously adopted thresholds are typically $1\sigma - 2\sigma$ above the mean background. For example, \cite{Ragan2014} uses a $^{13}$CO emission intensity threshold of $\sim 1~\rm K~km~s^{-1}$, while \cite{Abreu-Vicente2016} adopts a similar value, corresponding to $N(\rm H_2) \simeq 0.87 - 1.3 \times 10^{21}~\rm cm^{-2}$. The Herschel filaments identified by \cite{Wang2015} corresponds to structures denser than $N(\rm H_2) > 10^{21}~\rm cm^{-2}$.
To consistently identify filamentary structures across the simulated galaxy, we adopt a uniform threshold of $N_\mathrm{H_2, CO} = 1.0 \times 10^{21}~\rm cm^{-2}$. In addition, we exclude the central region within a galactocentric radius of $R_\mathrm{gal} \leq 2.5~\rm kpc$, as GMFs in previous observational catalogs are not located near the Galactic center. \textsc{FilFinder} is then used to reduce the isolated contours in the mask to single-pixel-wide skeletons representing filament spines. To remove spurious features caused by pixelization, we prune skeletons and branches shorter than 10 pc. These thresholds are motivated by two considerations: (1) the minimum GMF length observed in previous studies is $\sim 10~\rm pc$ (see Figure 10 of \citealt{Zucker2018}); (2) structures smaller than 10 times our spatial resolution may not be fully resolved.

\begin{figure*}
\begin{center}
\includegraphics[width=\textwidth]{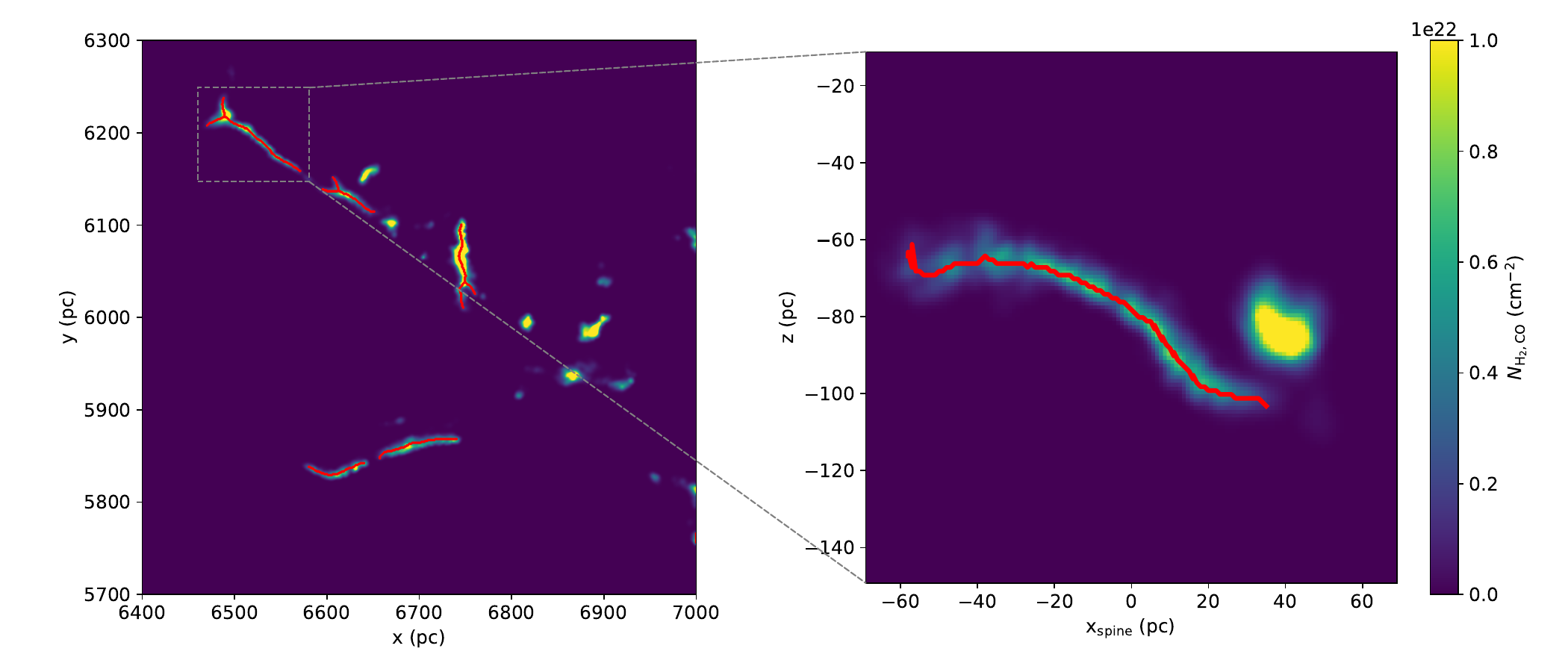}
\caption{Left panel: Face-on $N_\mathrm{H_2, CO}$ column density map of a $(600~\rm pc)^2$ region in the simulated galaxy. The spines of identified face-on GMFs are shown as red lines. Right panel: Edge-on view of the selected GMF in the grey box in the left panel. Note that the connected filament in the edge-on view consists of two components in the face-on view.}
\label{fig: 2 faceon edgeon fil}
\end{center}
\end{figure*}

For each identified filament after pruning, we measure four key physical properties: length $l$, width $w$, mass $M_\mathrm{CO}$, and aspect ratio $ar$. The filament length is defined as the length of the longest path along the skeleton (the filament spine). Filament width is determined using the \textsc{RadFil} package \citep{Zucker2018_RadFil}, which performs perpendicular cuts across the spine, measures the length of each cut from one side of the filament contour to the other, and then takes the median value as the filament width. Filament mass is calculated as the sum of molecular mass inferred from the synthetic CO emission within the filament contour. We adopt three criteria for a filament to be classified as a GMF: $l \geq 10~\rm pc$, $ar \geq 5$, and $M_\mathrm{CO} \geq 10^3~M_\odot$. The minimum aspect ratio follows \cite{Andr2014}, while the minimum mass ensures that each GMF is resolved by at least $\sim 10$ gas particles. Applying these criteria yields 1042 GMFs in the face-on synthetic observation, forming our face-on synthetic GMF catalog. The spines of these GMFs are illustrated as red lines over a $(600~\rm pc)^2$ region of the $N_\mathrm{H_2, CO}$ column density map in the left panel of \autoref{fig: 2 faceon edgeon fil}.

To generate our in-plane (edge-on) views of these GMFs, each face-on GMF is projected along its minor axis, determined from the eigenvector corresponding to the smaller eigenvalue of the covariance matrix of the filament contour. The edge-on projection considers only gas particles within the face-on GMF contour to exclude foreground and background gas. We analyze the impact of projection effects on GMF physical property measurements in \autoref{appendix}. The projection is centered on the CO-traced molecular mass of the selected gas particles. For the edge-on column density map, the window size is the spine length plus 10 pc to avoid boundary effects, the north vector is aligned with the simulation $z$-axis (perpendicular to the galactic plane), and the resolution is 1 pc. The edge-on view of a selected GMF is shown in the right panel of \autoref{fig: 2 faceon edgeon fil}. GMFs in the edge-on maps are identified using the same procedure as for the face-on maps, yielding a catalog of 742 GMFs. The reduced number of GMFs arises because some face-on GMFs consist of multiple overlapping components that are merged along a single line of sight, and individual components may not satisfy the identification criteria in the edge-on view. Using synthetic observations from two perpendicular lines of sight ensures that our edge-on GMFs are spatially connected structures. We illustrate the spatial distribution of these GMFs in the simulated galaxy as the cyan contours in the upper right panel of \autoref{fig: 1 galactic maps}.

\section{Results: Synthetic GMF catalog}
\label{sec:resultsA}

To assess whether the GMFs identified in our synthetic observations resemble those in previous catalogs, we compare four key physical properties of our edge-on synthetic GMF catalog to the observed values: length, width, mass, and linear mass. Length, width, and mass for each synthetic GMF were determined in \autoref{sec:GMF_identify}, while linear mass is calculated as the ratio of GMF mass to length.  

\begin{figure*}
\begin{center}
\includegraphics[width=\textwidth]{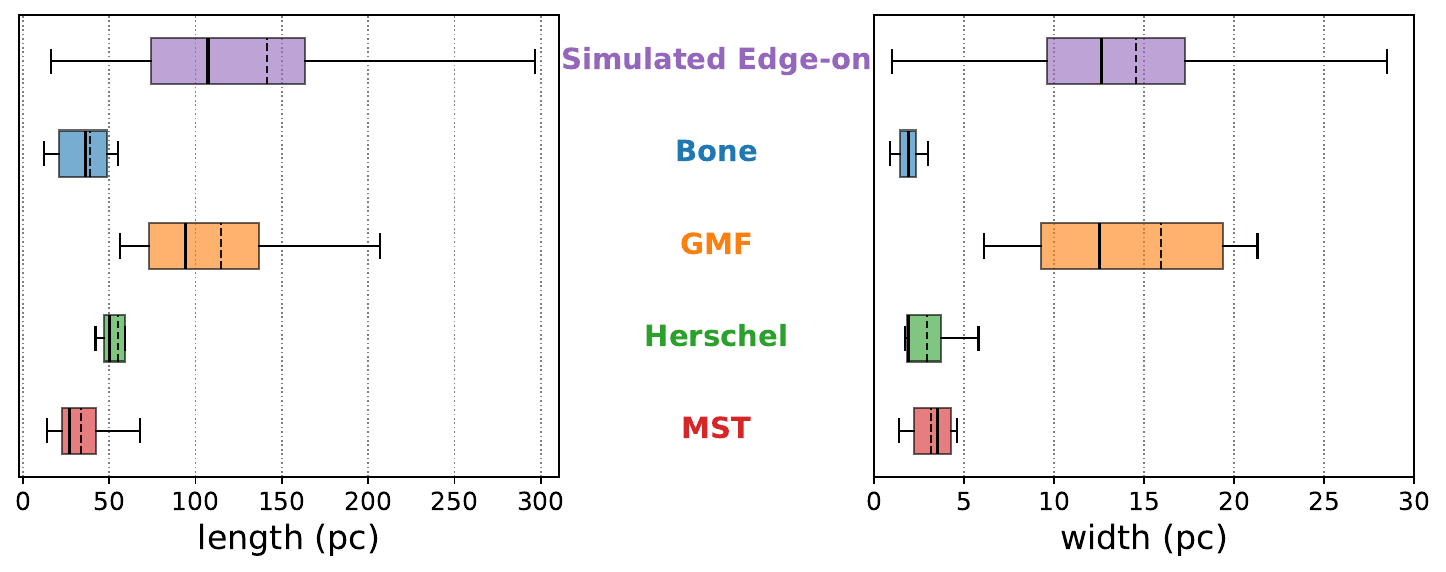}
\includegraphics[width=\textwidth]{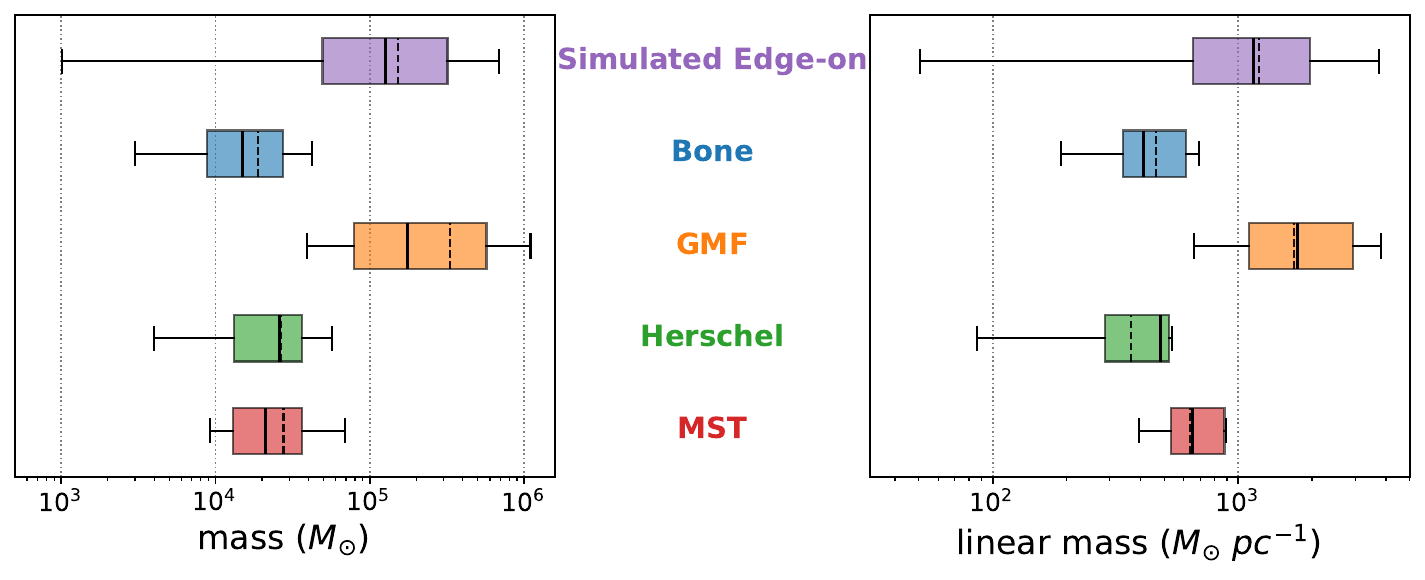}
\caption{Box-and-whisker plots showing the distributions of GMF length (top left), width (top right), mass (bottom left), and linear mass (bottom right). Solid lines indicate the median, while dashed lines indicate the mean. Colors and labels in the center identify the data source of each GMF sample.}
\label{fig: 3 length width mass linear mass}
\end{center}
\end{figure*}

For observed Galactic GMFs, \cite{Zucker2018} classified them into four categories based on filament selection criteria and summarized their physical properties. The categories include Milky Way Bone, Giant Molecular Filament, Large-scale Herschel, and Minimum-Spanning-Tree Bone, abbreviated as ``Bone", ``GMF", ``Herschel", and ``MST". We present the distributions of length, width, mass, and linear mass as box-and-whisker plots in \autoref{fig: 3 length width mass linear mass}. From the plot, we find that our synthetic GMFs closely reproduce the physical properties of the ``GMF" category, which corresponds to the largest molecular structures observed in the Galaxy \citep{Ragan2014, Abreu-Vicente2016}. This agreement arises from two main factors: (1) filament widths in other categories are only around a few parsecs, which are close to the 1 pc spatial resolution of our synthetic observations; (2) the universal column density threshold we adopt is based on the observations of the ``GMF" category. The effects of alternative filament identification algorithms are discussed in \autoref{sec: 5.1 GMF identification algorithm}.

We also summarize additional physical properties of our synthetic edge-on GMFs in \autoref{tab:edgeon_properties}, including mean column density $\bar{N}_\mathrm{H_2, CO}$, line-of-sight velocity gradient $\Delta v_\mathrm{LOS}/\Delta l$, velocity dispersion $\sigma_v$, and mean magnetic field strength $B$. The mean column density $\bar{N}_\mathrm{H_2, CO}$ is obtained from the total molecular mass traced by CO emission within the filament mask divided by the mask area. The line-of-sight velocity gradient is calculated as the ratio of the maximum $v_\mathrm{LOS}$ difference among all pixels inside the filament mask to the filament spine length $l$. Filament velocity dispersion\footnote{Note that the velocity dispersion here is the intrinsic dispersion derived from 3D velocity vectors. The result indicates the upper limit of observed velocity dispersion, which is calculated from LOS 1D components.} is computed as
\begin{equation}
\sigma_v = \sqrt{\sigma_{v,x}^2 + \sigma_{v,y}^2 + \sigma_{v,z}^2},
\end{equation}
where $\sigma_{v,i}$ ($i \in \{x, y, z\}$) is the CO-emission-weighted velocity dispersion of all gas particles within the filament mask along the $i$-axis. For the magnetic field, we first determine the magnitude of the field for all gas particles within the filament mask and then take the CO-emission-weighted mean. We do not perform a comparison to observations for these properties, as they are not available in all GMF catalogs.

\begin{table*}
\centering
\begin{tabular}{lcccc}
\hline
 & $N_{\rm H_2, CO}$ ($\times$ $10^{21}$ cm$^{-2}$) & $\Delta v_\mathrm{LOS}/\Delta l$  (km s$^{-1}$ pc$^{-1}$) & $\sigma_v$ (km s$^{-1}$) & $B$ ($\mu$G) \\
\hline
Mean & 5.75 & 0.048 & 5.31 & 15.88 \\
25th percentile & 3.81 & 0.028 & 3.05 & 10.85 \\
50th percentile & 5.10 & 0.040 & 4.48 & 14.58 \\
75th percentile & 6.89 & 0.055 & 6.50 & 19.13 \\
\hline
\end{tabular}
\caption{Mean values and percentiles of edge-on synthetic GMF properties.}
\label{tab:edgeon_properties}
\end{table*}

The mean value of the synthetic $N_{\rm H_2, CO}$ column density is similar to the typical value of $4.8\times10^{21}~\rm cm^{-2}$ observed for the ``GMF" category. The magnetic field strength (10–20 $\mu$G) is also consistent with values measured in GMCs \citep{Crutcher2012_BFieldReview}. Regarding the velocity dispersion, adopting the mean length $l \approx 140~\rm pc$ and mean width $w \approx 16~\rm pc$ of the edge-on GMFs gives a characteristic size $L = \sqrt{lw} \approx 47~\rm pc$. Using Larson's Law, $\sigma_v~(\rm km~s^{-1}) = 1.10 L~(\rm pc)^{0.38}$ \citep{Larson1981}, such a size corresponds to $\sigma_v \approx 4.8~\rm km~s^{-1}$, which lies between the median and mean velocity dispersions of our synthetic GMFs. Line-of-sight velocity gradients have not been systematically studied for $\sim 100$ pc-long GMFs, but for comparison, \cite{Goodman2014} measured a gradient of $0.025~\rm km~s^{-1}~pc^{-1}$ for the ``Nessie" filament, comparable to the values found in our synthetic catalog. Therefore, our synthetic GMFs are consistent with real observations when considering these additional physical properties. \cite{Wang2016} and \cite{Ge2022} report median values of $0.33~\rm km~s^{-1}~pc^{-1}$ and $0.47~\rm km~s^{-1}~pc^{-1}$, respectively, for the velocity gradients of their ``MST" GMF catalogs, which are significantly larger than the result from our synthetic catalog. Such discrepancy may originate from the velocity difference criteria applied in the MST algorithm as explained in \autoref{sec: 5.1 GMF identification algorithm}.

\begin{figure}
    \centering
    \includegraphics[width=\linewidth]{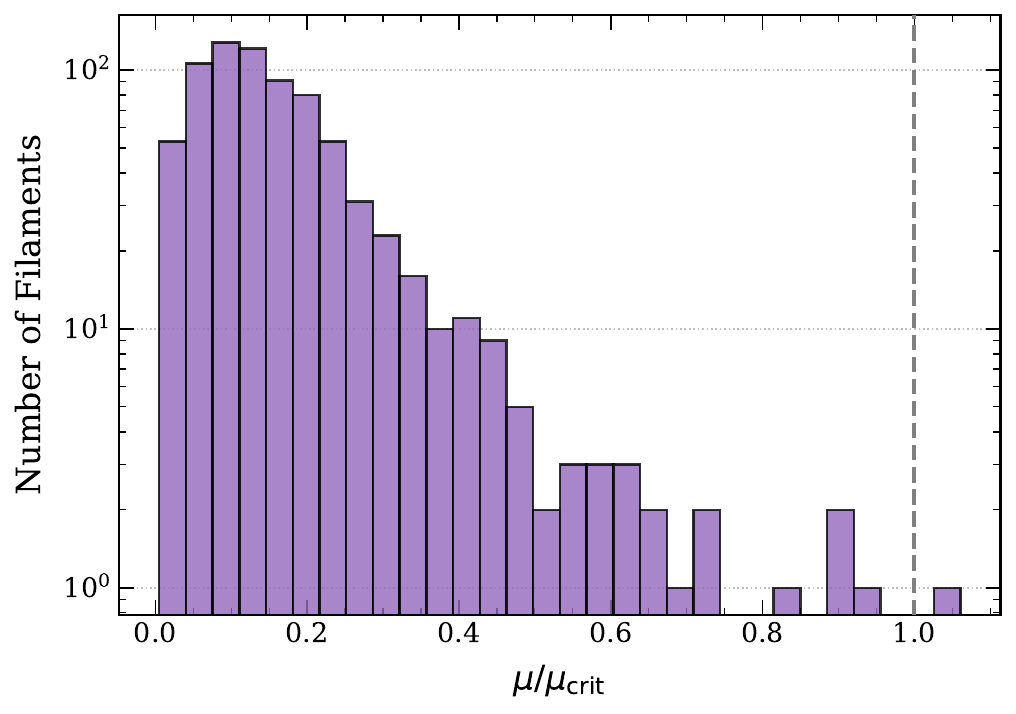}
    \caption{Histogram of the ratio between the synthetic GMF linear mass $\mu$ and the virial linear mass $\mu_\mathrm{vir}$. The vertical grey dashed line indicates $\mu/\mu_\mathrm{vir} = 1$.}
    \label{fig: 5 linear mass ratio}
\end{figure}

We also perform a one-dimensional (1D) virial analysis on the synthetic GMFs. As shown in \autoref{tab:edgeon_properties}, the velocity dispersions of the GMFs are much larger than the thermal velocity dispersion $c_s \sim 0.3~\rm km~s^{-1}$ for $T \approx 20$ K, indicating that turbulent motions dominate their gravitational stability. The synthetic GMF virial line mass is computed as
\begin{equation}
\mu_\mathrm{vir} = \frac{2\sigma_v^2}{G} \sim \left( \frac{\sigma_v}{1~\rm km~s^{-1}} \right)^2~\rm M_\odot~pc^{-1},
\label{eq: virial linear mass}
\end{equation}
where $G$ is the gravitational constant. In \autoref{fig: 5 linear mass ratio}, we show a histogram of the ratio between the GMF linear mass $\mu$ and the virial line mass $\mu_\mathrm{vir}$. The figure demonstrates that nearly all synthetic GMFs are gravitationally subcritical.

\section{Results: GMF life cycle}
\label{sec:resultsB}

After identifying the synthetic GMFs, we trace their evolution across different simulation snapshots. The typical life cycle of an isolated GMF can be divided into three stages. The first is the \textit{formation stage}, during which GMFs assemble from the accretion of atomic gas. The second is the \textit{fragmentation stage}, when turbulent perturbations cause the GMFs to break into dense clumps, triggering rapid star formation. The third is the \textit{dispersion stage}, where stellar feedback—primarily from massive stars—disperses the molecular gas. Owing to their large spatial extent, GMFs are influenced by physical processes acting across multiple scales. In the following, we discuss the roles of the dominant mechanisms during each stage of the GMF life cycle.

\subsection{Formation mechanism}
\subsubsection{Dominant mechanism: magnetized turbulence}
\label{sec: dominant mechanism}
\begin{figure} 
\centering 
\includegraphics[width=\linewidth]{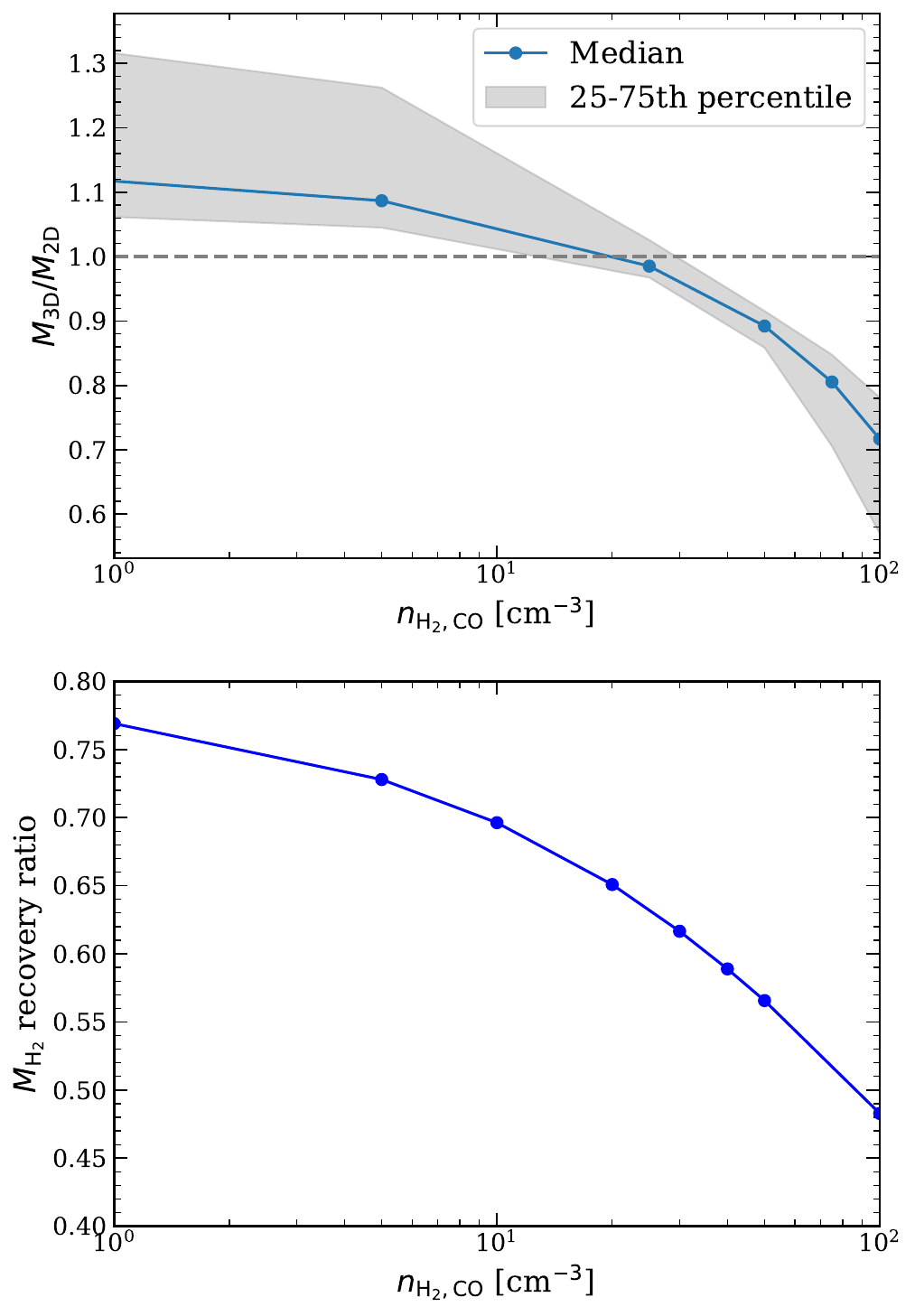} 
\caption{Top panel: Ratio between the molecular mass from 3D contours surrounding the CoM of GMF $M_\mathrm{3D}$ and the molecular mass from 2D synthetic observations $M_\mathrm{2D}$. The x-axis shows the value of the number density threshold of different 3D contours. The blue dotted line shows the median value of the ratios from the synthetic edge-on GMF catalog; while the grey band illustrates the 25th to 75th range of the ratios. Bottom panel: ratio between the molecular hydrogen mass above a given $n_\mathrm{H_2,CO}$ threshold and the total molecular hydrogen mass.} 
\label{fig: 6 fil3D threshold} 
\end{figure}

As discussed in Section 5 of \cite{Hacar2022}, turbulence, magnetic fields, stellar feedback, and galactic dynamics all contribute to filament formation. The one-dimensional analysis in \autoref{fig: 5 linear mass ratio} indicates that turbulence dominates over gravity for most GMFs. However, to fully assess the relative importance of different physical mechanisms, a three-dimensional virial analysis is required, both for the GMFs themselves and for the surrounding atomic regions where gas accretion occurs. This requires identifying the 3D counterparts of the 2D GMFs seen in synthetic observations. For each GMF, we first construct a cubic box with a side length 20 pc larger than the GMF length and interpolate the gas properties of all particles onto a 1 pc voxel grid. We then draw 3D molecular density contours centered on the GMF center of mass (CoM) with thresholds $n_\mathrm{H_2,CO} \in \{0.5, 5, 25, 50, 75, 100\} \; \rm cm^{-3}$, where $n_\mathrm{H_2,CO}$ is the CO-derived molecular hydrogen number density. From these contours, we calculate the enclosed molecular mass $M_\mathrm{3D}$ and compare it with the 2D mass estimate $M_\mathrm{2D}$. The top panel of \autoref{fig: 6 fil3D threshold} shows the median and interquartile range of $M_\mathrm{3D}/M_\mathrm{2D}$ as a function of threshold. We find that $M_\mathrm{3D}/M_\mathrm{2D} \sim 1$ at $n_\mathrm{H_2,CO} = 20 \; \rm cm^{-3}$. We therefore adopt $n_\mathrm{H_2,CO} = 20 \; \rm cm^{-3}$ as the fiducial threshold for identifying GMFs in 3D, as it ensures the consistency between the masses measured from 2D projections and 3D structures.

Although this threshold is much lower than both the mean density of the synthetic GMFs and the typical density of CO-bright gas ($\bar{n}_\mathrm{H_2,CO} \approx 100 \; \mathrm{cm^{-3}}$, \citealp{Bolatto2013}), it corresponds to the diffuse molecular–gas regime ($\sim 30 \; \mathrm{cm^{-3}}$) observed in the Milky Way \citep{Klessen2016}. Moreover, this density is approximately where the molecular gas begins to transition from the CO-dark to the CO-bright phase in our synthetic observations. To illustrate this, we vary the $n_\mathrm{H_2,CO}$ threshold from $1 \; \mathrm{cm^{-3}}$ to $100 \; \mathrm{cm^{-3}}$, compute the ratio between the molecular hydrogen mass above each threshold and the total molecular hydrogen mass, and plot these ratios in the bottom panel of \autoref{fig: 6 fil3D threshold}. At the adopted threshold of $n_\mathrm{H_2,CO} = 20 \; \mathrm{cm^{-3}}$, approximately $65\%$ of the molecular mass is recovered, which is broadly consistent with the observed range ($50\%$–$67\%$) in the solar neighborhood \citep{Grenier2005}. Varying the threshold by a factor of two around $20 \; \mathrm{cm^{-3}}$ changes the total mass of the identified 3D GMFs by only $\sim 10\%$, demonstrating the robustness of our subsequent analysis.

\begin{figure*}
\begin{center}
\includegraphics[width=\textwidth]{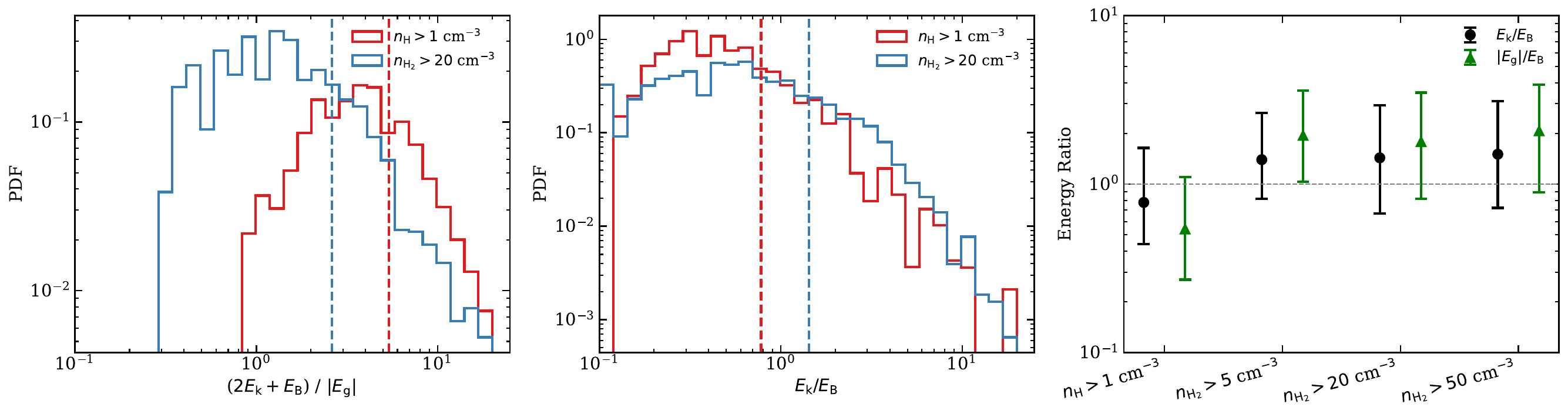}
\caption{Left panel: PDFs of $(2E_k +E_B)/|E_g|$. The red histogram shows the PDF from 3D number density contours, corresponding to the atomic gas region containing GMF. The blue histogram show the PDF from GMF itself. The vertical lines represent the median values of the histograms with the same color. Middle panel: same as the left panel, but showing the PDFs of $E_k/E_B$. Right panel: Median energy ratios versus density threshold. Black circles and green triangles represent $E_{\rm k}/E_{\rm B}$ and $|E_{\rm g}|/E_{\rm B}$, respectively, with error bars showing the interquartile range (25th-75th percentiles).}
\label{fig: 7 energy ratios}
\end{center}
\end{figure*}

After linking 2D and 3D GMFs, we compute their gravitational potential energy $E_g$, magnetic energy $E_B$, and turbulent energy $E_k$ using the 3D voxel grid. The gravitational potential $\phi$ is obtained with the adaptive \textsc{pytreegrav} library \citep{Grudic2021_pytreegrav}, yielding $E_g = \frac{1}{2}\sum \rho_i \phi_i$. The magnetic energy is given by $E_B = \frac{1}{8\pi}\sum B_i^2 V_i$, where $V_i = 1 \; \rm pc^3$ is the voxel volume. The turbulent energy $E_k$ is determined from the velocity dispersion relative to the mass-weighted mean velocity: $E_k = \frac{1}{2}\sum \rho_i \left[(v_{x,i}-\bar{v}_x)^2 + (v_{y,i}-\bar{v}_y)^2 + (v_{z,i}-\bar{v}_z)^2\right]$. These energies are calculated both within the GMF masks ($n_\mathrm{H_2,CO} > 20 \; \rm cm^{-3}$) and including the surrounding atomic regions ($n_\mathrm{H} > 1 \; \rm cm^{-3}$). The histograms in the left and middle panels of \autoref{fig: 7 energy ratios} show the probability distribution functions (PDFs) of $(2E_k +E_B)/|E_g|$ and $E_k/E_B$. We find $(2E_k +E_B)/|E_g| \gg 1$ on both scales, indicating that magnetized turbulence dominates over gravity during mass accretion. Meanwhile, $E_k/E_B \sim 1$, suggesting comparable contributions from turbulence and magnetic fields.

\subsubsection{Role of magnetic field}

In order to study the relative importance of magnetic field across different density regimes, we calculate $E_k/E_B$ and $|E_g|/E_B$ in 3D contours with 4 density levels: $n_{\rm H} > 1$ cm$^{-3}$, $n_{\rm H_2} > 5$ cm$^{-3}$, $n_{\rm H_2} > 20$ cm$^{-3}$, and $n_{\rm H_2} > 50$ cm$^{-3}$. Then we plot the 25th, 50th, and 75th percentiles of these energy ratios in the right panel of \autoref{fig: 7 energy ratios}. The results show that in the atomic regime ($n_{\rm H} > 1$ cm$^{-3}$), the magnetic energy is comparable to the kinetic energy and dominant over gravitational energy, but become less important than the other two energy sources when molecules start to form.

\begin{figure*}
\begin{center}
\includegraphics[width=\textwidth]{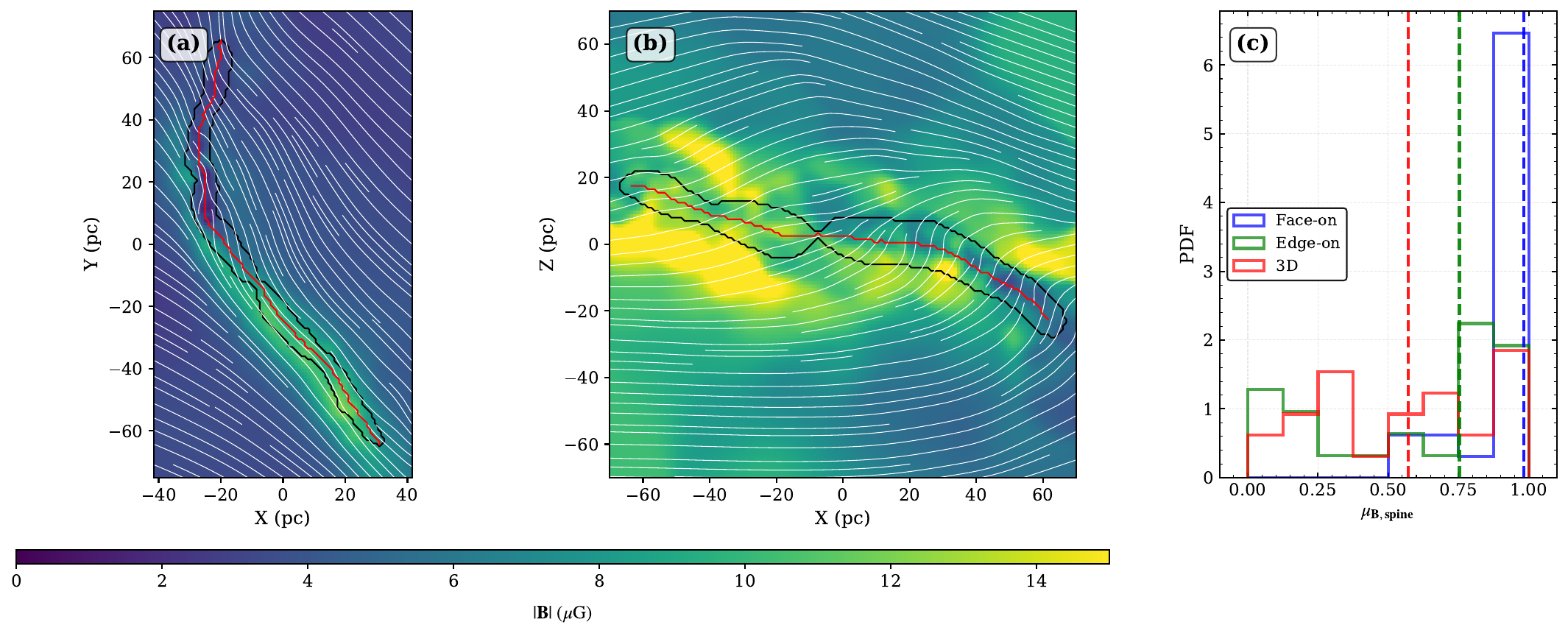}
\caption{\textbf{(a)} Face-on view of the magnetic field strength $|\mathbf{B}|$ surrounding an example GMF with magnetic field streamlines (white lines) overlaid. The filament spine is indicated by the red line, and the GMF boundary is outlined in black.
\textbf{(b)} Same as (a), but showing the edge-on view projected along the GMF's minor axis.
\textbf{(c)} PDFs of the cosine of the angle between the magnetic field vector and the filament spine, $\mu_{\mathbf{B},\mathbf{spine}}$, for face-on view (blue), edge-on view (green), and 3D space (red). The vertical dashed lines indicate the median values for each distribution.}
\label{fig: example B field map}
\end{center}
\end{figure*}

To further evaluate the role of magnetic fields in GMF evolution, we examine their alignment with gas structures. We illustrate the mass-weighted B-field mean strength projection map surrounding an example GMF from two perspectives: a face-on view (panel a) and an edge-on view (panel b) in \autoref{fig: example B field map}. Observational studies of 2D projected fields typically employ the histogram of relative orientations (HRO), the alignment measure (AM) parameter, or the projected Rayleigh statistic (PRS) \citep{Soler2013, Gonzalez-Casanova2017_AM, Yuen2017_AM, Jow2018_PRS} for this purpose, but since we have access to the full 3D structure in our simulations we opt to measure the degree of alignment between a pair of vector fields $\mathbf{v}_1$ and $\mathbf{v}_2$ directly by computing the distribution of the absolute value of cosine of the angle between them, $\mu_{\mathbf{v}_1,\mathbf{v}_2} = |\mathbf{v}_1 \cdot \mathbf{v}_2| / v_1 v_2$, where $v_1 = |\mathbf{v}_1|$ and similarly for $v_2$. If the two vector fields are randomly oriented, the PDF of $\mu$ is expected to be uniform between 0 and 1. In contrast, a median value of the PDF close to 0 indicates a preference for perpendicular orientations, while a value close to 1 indicates a preference for parallel orientations.

To justify our choice of 3D over 2D analysis, we quantify the magnetic field-to-spine alignment, $\mu_{\mathbf{B},\mathbf{spine}}$, for the example GMF in \autoref{fig: example B field map}. We compute this metric from three perspectives: face-on, edge-on, and in 3D space. The spines for the projected views are identified using \textsc{FilFinder} (as in \autoref{sec:GMF_identify}), while the 3D spine is derived from the mask in \autoref{sec: dominant mechanism} using \textsc{FilFinderPPP}, the 3D extension of \textsc{FilFinder}. To account for variations along the filament, we divided each spine into 5-pc-long segments. For each segment, we calculated the mean spine direction from its end points and the mean magnetic field direction from the average of B-field vectors within a 5-pc radius. We then computed $\mu_{\mathbf{B},\mathbf{spine}}$ for all segments in the three cases.

The resulting PDFs and median values are shown in panel (c) of \autoref{fig: example B field map}. The face-on view indicates a nearly parallel alignment, whereas the edge-on view shows a mix of parallel and perpendicular alignments. In contrast, the 3D analysis reveals no preferred alignment. These results demonstrate that projection effects can introduce significant uncertainties in alignment analysis, underscoring the necessity of 3D information for accurately determining the role of magnetic fields.

\begin{figure*}
\begin{center}
\includegraphics[width=\textwidth]{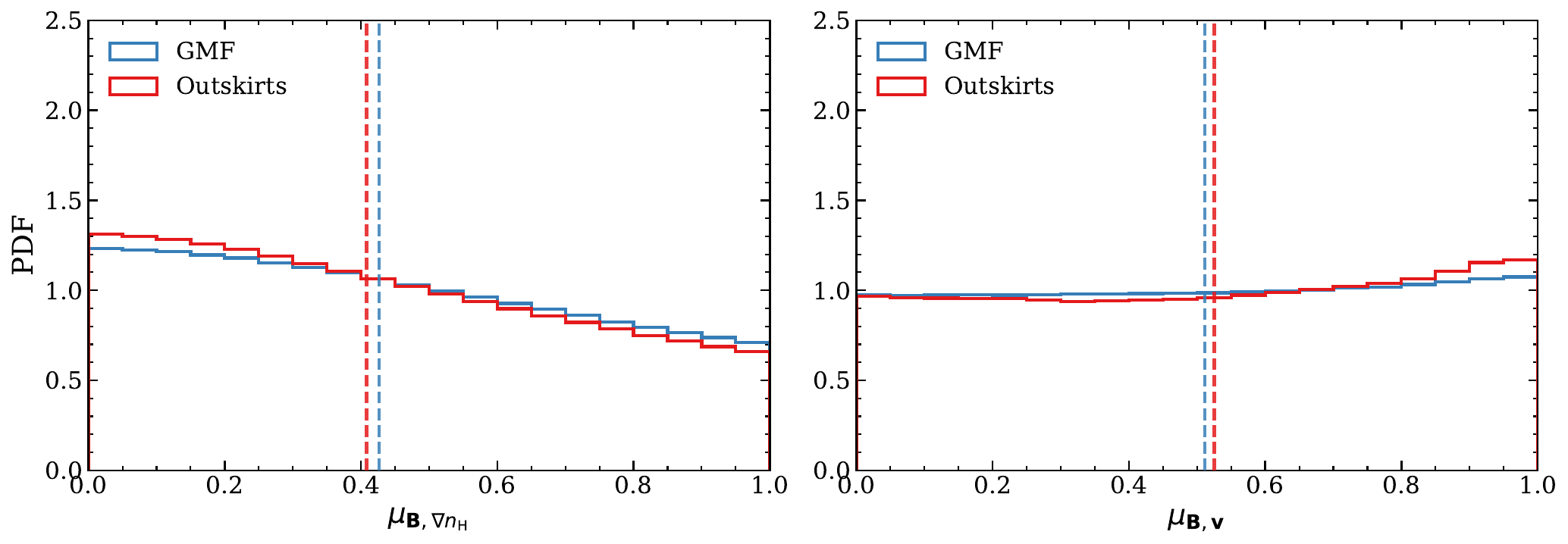}
\caption{Distribution of cosine angles between magnetic field vectors and (left) density gradients $\nabla n_{\rm H}$ and (right) relative velocity vectors $\mathbf{v}$. Blue and red histograms represent GMFs ($n_{\rm H_2} > 20$ cm$^{-3}$) and atomic outskirts ($n_{\rm H_2} < 20$ cm$^{-3}$, $n_{\rm H} > 1$ cm$^{-3}$), respectively. Dashed vertical lines indicate median values.}
\label{fig: 8 cos histogram}
\end{center}
\end{figure*}

To directly study the role of magnetic field in the dynamics of GMF, we first compute the distribution of cosine for the magnetic field and the density gradient, $\mu_{\mathbf{B}, \nabla n_\mathrm{H}}$, over all voxels both inside GMFs ($n_\mathrm{H_2, CO} \geq 20 \; \rm cm^{-3}$) and in the surrounding atomic outskirts ($n_\mathrm{H_2, CO} < 20 \; \rm cm^{-3}$, $n_\mathrm{H} > 1 \; \rm cm^{-3}$). We plot the results in the left panel of \autoref{fig: 8 cos histogram}. The distributions show a very weak preference for perpendicular alignment in both regimes. We also compute the distribution of cosine angles between the magnetic field and relative gas velocity vectors, $\mu_{\mathbf{B},\mathbf{v}}$, which we show in the right panel of \autoref{fig: 8 cos histogram}. Here we define the relative gas velocity of each voxel as the velocity at that position minus the mass-weighted mean velocity of the GMF. We find no significant evidence for a preferred alignment in either the GMFs or their atomic outskirts. 

These results suggest that magnetic fields do not strongly regulate gas flow directions, particularly in the accreting envelopes of GMFs. Instead, similar to turbulence, their main role during GMF formation is to provide support against gravity rather than to channel gas inflows.

\subsubsection{Source of turbulent energy}
Since magnetized turbulence is the main driving mechanism during the GMF formation phase, we now examine the physical processes that inject such turbulent energy. Possible contributors include gravitational instabilities, galactic rotation, and stellar feedback \citep{Elmegreen2004_turbulenceReview}. Given the large spatial scale of the simulated GMFs ($\sim100$ pc), galactic shear is expected to dominate.  

\begin{figure*}
\begin{center}
\includegraphics[width=\textwidth]{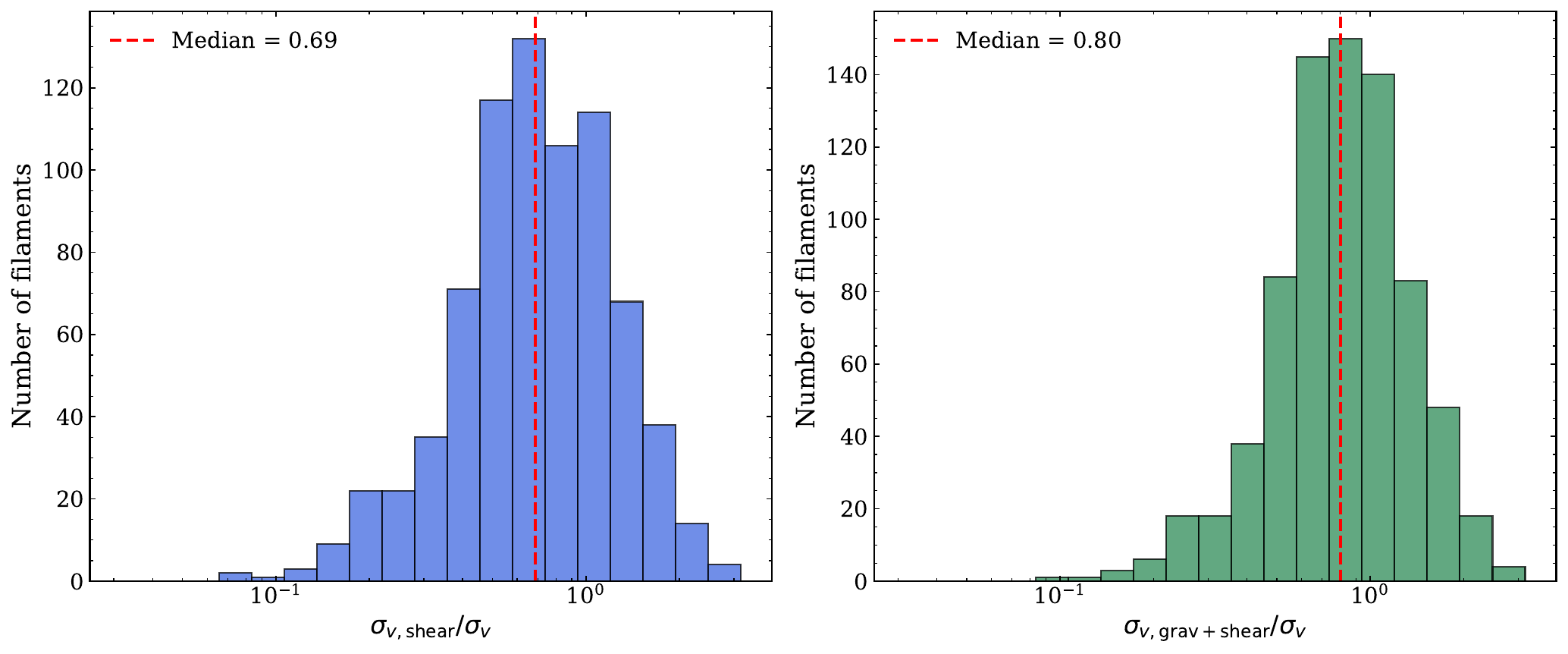}
\caption{Left: histogram of the ratios between shear-induced velocity dispersion and total velocity dispersion. Right: same as left, but including contributions from both shear and self-gravity. The red dashed line marks the median ratio.}
\label{fig: 9 shear contribution}
\end{center}
\end{figure*}

To quantify this, we first derive the angular velocity profile $\Omega(r)$ by averaging over radial bins of 50 pc. For each edge-on GMF we measure its projected radial extent $l_r$, and following \cite{Xie2025_shearSigmaV}, define the shear-induced velocity dispersion as  
\begin{equation}
    \sigma_{v, \rm shear} = \kappa l_r = rl_r \left| \frac{d\Omega(r)}{dr} \right|,
\end{equation}
where $\kappa$ is the local shear rate and $r$ is the galactocentric radius of the GMF’s center of mass. The histogram of $(\sigma_{v, \rm shear}/\sigma_{v})$ in the left panel of \autoref{fig: 9 shear contribution} shows that shear typically accounts for $\sim70\%$ of the total velocity dispersion.  

We next consider the contribution from self-gravity. As shown in \autoref{fig: 5 linear mass ratio} and \autoref{fig: 7 energy ratios}, most GMFs are sub-critical and therefore not undergoing global collapse. Instead, we estimate an upper limit to the gravitational contribution, defined as the dispersion required to maintain linear virial equilibrium:  
\begin{equation}
    \sigma_{v, \rm grav} = \left( \frac{GM}{2l} \right)^{1/2}.
\end{equation}
Combining both contributions gives  
\begin{equation}
    \sigma_{v, \rm grav+shear} = \left( \sigma_{v, \rm grav}^2 + \sigma_{v, \rm shear}^2 \right)^{1/2}.
\end{equation}
As shown in the right panel of \autoref{fig: 9 shear contribution}, the median ratio $(\sigma_{v, \rm grav+shear}/\sigma_{v})$ is $\sim80\%$, only $\sim10\%$ higher than the shear-only case. Thus, self-gravity plays a much smaller role in driving turbulence than galactic shear.  

\subsubsection{Role of stellar feedback}
The remaining $\sim20\%$ of the turbulent energy likely originates from stellar feedback, which can strongly reshape the gas distribution and promote filamentary structures due to its anisotropic nature. Directly quantifying this effect is difficult because feedback interacts non-locally with the galactic environment. Instead, we test its impact statistically by correlating the number of GMFs with feedback strength in the surrounding ISM.  

We divide the face-on galactic map (\autoref{fig: 1 galactic maps}) into square patches of side length 2.5 kpc, large enough to capture the local environment while yielding more than 30 regions for systematic comparison. For each patch we compute: (1) the total gas mass, (2) the number of GMFs with centers inside the patch, (3) the number of recent ($1$--$10$ Myr) supernovae, and (4) the total H$\alpha$ luminosity $L_\mathrm{H\alpha}$ tracing photoionization. To minimize the bias from uneven mass distribution, we normalize GMF counts, SNe, and $L_\mathrm{H\alpha}$ by the gas mass in each patch.  

\begin{figure*}
\begin{center}
\includegraphics[width=\textwidth]{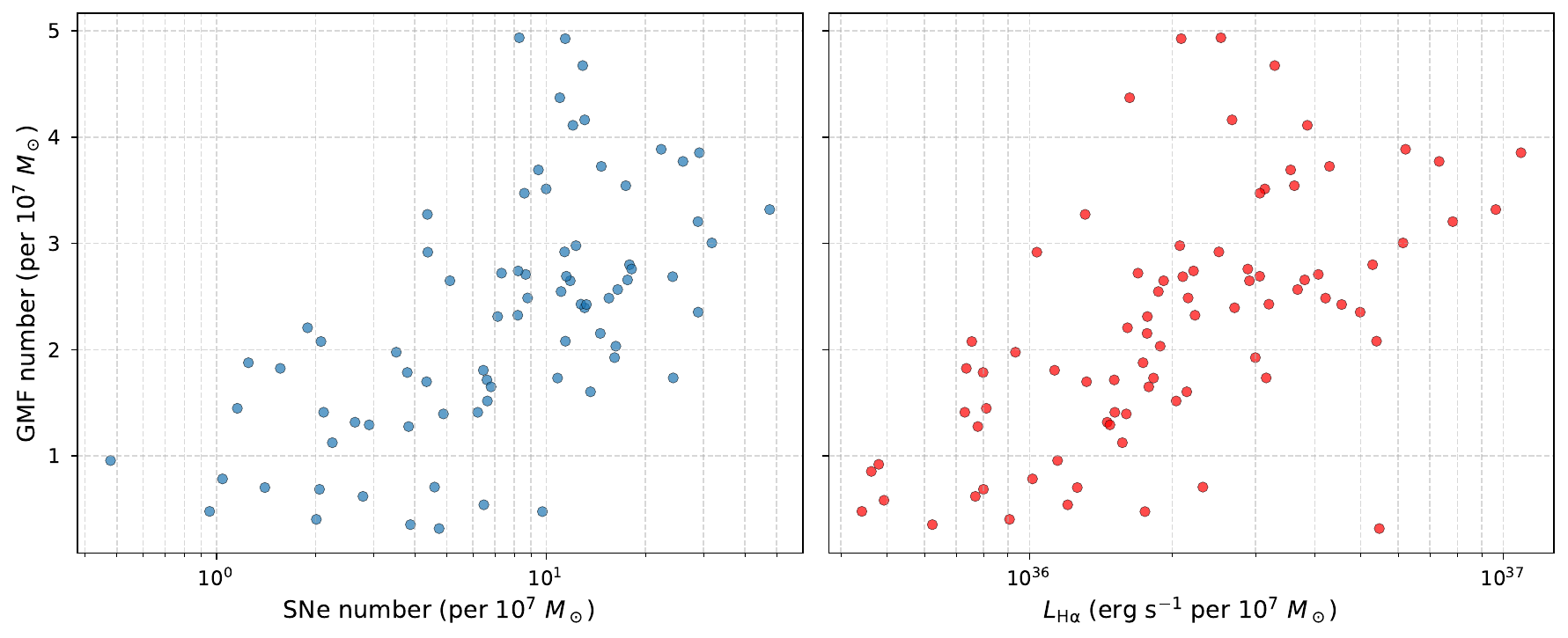}
\caption{Number of GMFs per unit gas mass in (2.5 kpc)$^2$ patches, plotted against the number of recent supernovae (left) and the total H$\alpha$ luminosity $L_\mathrm{H\alpha}$ (right). All variables are normalized by patch gas mass.}
\label{fig: 10 GMF vs stellar feedback}
\end{center}
\end{figure*}

The correlations are shown in \autoref{fig: 10 GMF vs stellar feedback}. We find that the number of GMFs increases with both the local SN rate and H$\alpha$ luminosity. The Spearman coefficients are $R_s = 0.64$ for GMF number vs. SNe and $R_s = 0.66$ for GMF number vs. $L_\mathrm{H\alpha}$, both indicating significant positive monotonic trends. These results suggest that stellar feedback from massive stars enhances the formation of GMFs. However, we cannot from this analysis alone rule out the possibility that the causation runs in the other direction -- that the presence of filaments causes (or shares a common cause with) star formation, which in turn leads to stronger feedback in locations where filaments are found. Additional simulations without stellar feedback physics are still needed to directly evaluate the role of stellar feedback in GMF formation.

\subsection{GMF evolution}
\label{sec: 4.2 GMF evolution}

\begin{figure}
    \centering
    \includegraphics[width=\linewidth]{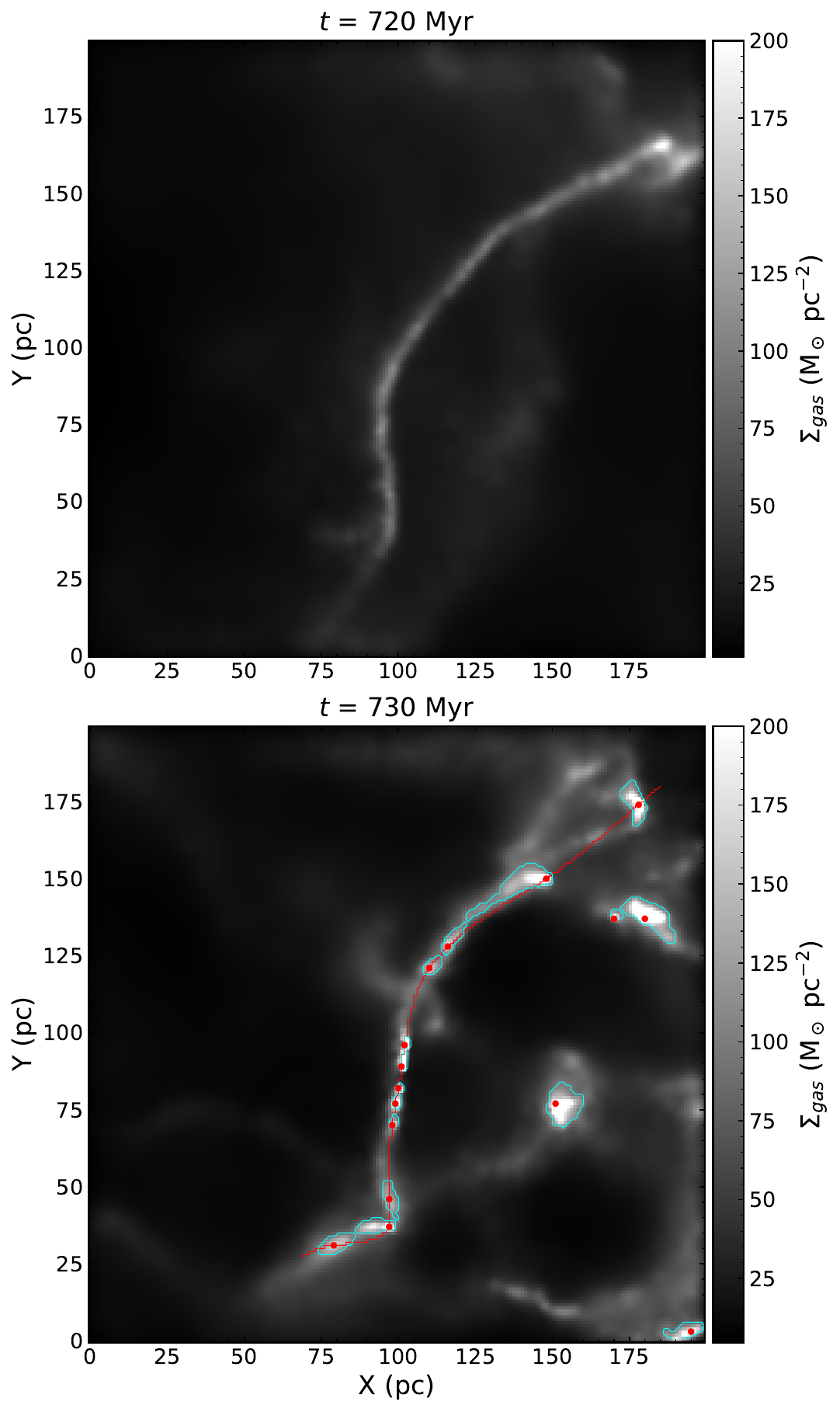}
    \caption{Gas surface density maps of a representative GMF at $t=720$ Myr (top) and $t=730$ Myr (bottom). In the later stage, cyan contours mark clumps identified using \textsc{astrodendro}; red dots indicate the local maxima of clumps; and red lines connect clumps with similar 3D velocities (differences $< 5$ km s$^{-1}$).}
    \label{fig: 11 Fragmentation example}
\end{figure}

When tracing the evolution of simulated GMFs, a common pathway is fragmentation into a number of dense clumps. \autoref{fig: 11 Fragmentation example} illustrates this process for one example GMF: at $t = 720$ Myr (top panel) the structure appears as a long, coherent filament, while by $t = 730$ Myr (bottom panel) it has fragmented into a dozen compact clumps distributed along the original filament spine.  

Two classical theoretical frameworks have been widely invoked to explain such fragmentation. The first is the spherical Jeans instability model \citep{Jeans1902}, which predicts local gravitational collapse if perturbations exceed the Jeans length:  
\begin{equation}
    l_\mathrm{J} = c_\mathrm{s}\sqrt{\frac{\pi}{G\bar{\rho}}} 
    = \sqrt{\frac{\pi k_\mathrm{B}T}{\mu m_\mathrm{p} G\bar{\rho}}},
\end{equation}
where $\mu = 2.33$ is the mean molecular weight of molecular gas and $m_\mathrm{p}$ is the proton mass. In this scenario, the mean clump separation after fragmentation is expected to be comparable to $l_\mathrm{J}$.  

A second framework is the ``sausage'' instability model, which considers the fragmentation of a self-gravitating fluid cylinder \citep{Chandrasekhar1953, Nagasawa1987}. In this case, the average spacing between clumps is set by the critical wavelength of the fastest-growing unstable mode. For an isothermal infinite gas cylinder with scale height $H$, the critical wavelength is  
\begin{equation}
    \lambda_\mathrm{crit} = 22H,
\end{equation}
where  
\begin{equation}
    H = \sqrt{\frac{c_\mathrm{eff}^2}{4\pi G\rho_\mathrm{c}}},
\end{equation}
$c_\mathrm{eff}$ is the effective sound speed and $\rho_\mathrm{c}$ is the central gas density of the filament \citep{Nagasawa1987, Inutsuka1992}.  

While useful for providing insight, this simplified model neglects magnetic fields and self-rotation of the filament. To incorporate these effects, \cite{Nakamura1993} and \cite{Hanawa1993} derived a more general expression, showing that the characteristic spacing is related to the filament structure itself:  
\begin{equation}
    \lambda_\mathrm{crit} = 5.14 \, \mathrm{FWHM},
\end{equation}
where FWHM is the full width at half maximum of the GMF column density profile.

In order to examine which theoretical model best describes the fragmentation process of the simulated GMFs, we trace the evolution of selected GMFs to the fragmentation stage, determine the average spacings between clumps, and compare the values to those predicted by theoretical models. Such analysis is a multi-step process that combines spatial clustering, curve fitting, and spectral analysis. The first step is to select GMFs with $ar > 10$ and no significant branches (branch lengths larger than 10 pc). The reason is that clumps formed on a single-branch, very elongated structure make it easier for later average spacing calculation and model testing. After selection, we have 35 GMFs remaining. The second step is to trace the evolution of each GMF by plotting square gas surface density $\Sigma_\mathrm{gas}$ maps for each snapshot since $t = 720$ Myr. Each map is centered on the CoM of the GMF with a side length 20 pc longer than the GMF length and a pixel size of 1 pc. We choose true $\Sigma_\mathrm{gas}$ maps over synthetic CO observations to more precisely capture the underlying physical mechanisms. Then we use \textsc{astrodendro} \citep{Robitaille2019_astrodendro} to find dense clumps from the $\Sigma_\mathrm{gas}$ maps. Our GMF identification threshold $N_\mathrm{H_2, CO} = 1.0 \times 10^{21}$ cm$^{-2}$ corresponds to a gas surface density $\Sigma_\mathrm{gas, min} = 20 \, M_\odot \, \rm pc^{-2}$ with a hydrogen mass ratio of 0.76. Therefore, for consistent comparison across GMFs, when computing the dendrogram we set the \textit{min\_value} (clump threshold) to $5\Sigma_\mathrm{gas, min} = 100 \, M_\odot \, \rm pc^{-2}$, \textit{min\_delta} (clump minimum height) to $\Sigma_\mathrm{gas, min}$, and \textit{min\_npix} (clump minimum size) to 5 pixels. We find 21 GMFs out of the 35 selected ones fragment into at least three clumps during the evolution tracing. Other GMFs stop fragmenting mainly due to environmental disturbances, such as cloud–cloud collisions. These 21 GMFs form our sub-sample for theoretical model testing.

For each GMF in the sub-sample, we choose the snapshot with the most identified clumps, then group the clumps by velocity similarity using hierarchical clustering with a maximum 3D velocity difference of 5 km s$^{-1}$. The most populous velocity group is selected, and its constituent clumps are connected via a Minimum Spanning Tree (MST) algorithm to establish their spatial ordering along the filament's principal axis. A cubic spline is then fitted through the ordered clump positions, extended by 10 pc at both ends to capture the full filament extent. We plot the identified clumps and the spline fitted from the example GMF as cyan contours and the red line in the bottom panel of \autoref{fig: 11 Fragmentation example}, respectively. The surface density is extracted along this spline curve and smoothed using a Gaussian kernel with a 3-pc standard deviation to reduce noise. Finally, the Lomb–Scargle periodogram analysis is applied to the smoothed density profile, with a maximum period of half the curve length and a minimum period of the smoothing kernel size. The dominant period is identified as the wavelength corresponding to the peak power in the resulting power spectrum, with statistical significance assessed using a 75\% false alarm probability threshold. 

\begin{figure}
    \centering
    \includegraphics[width=\linewidth]{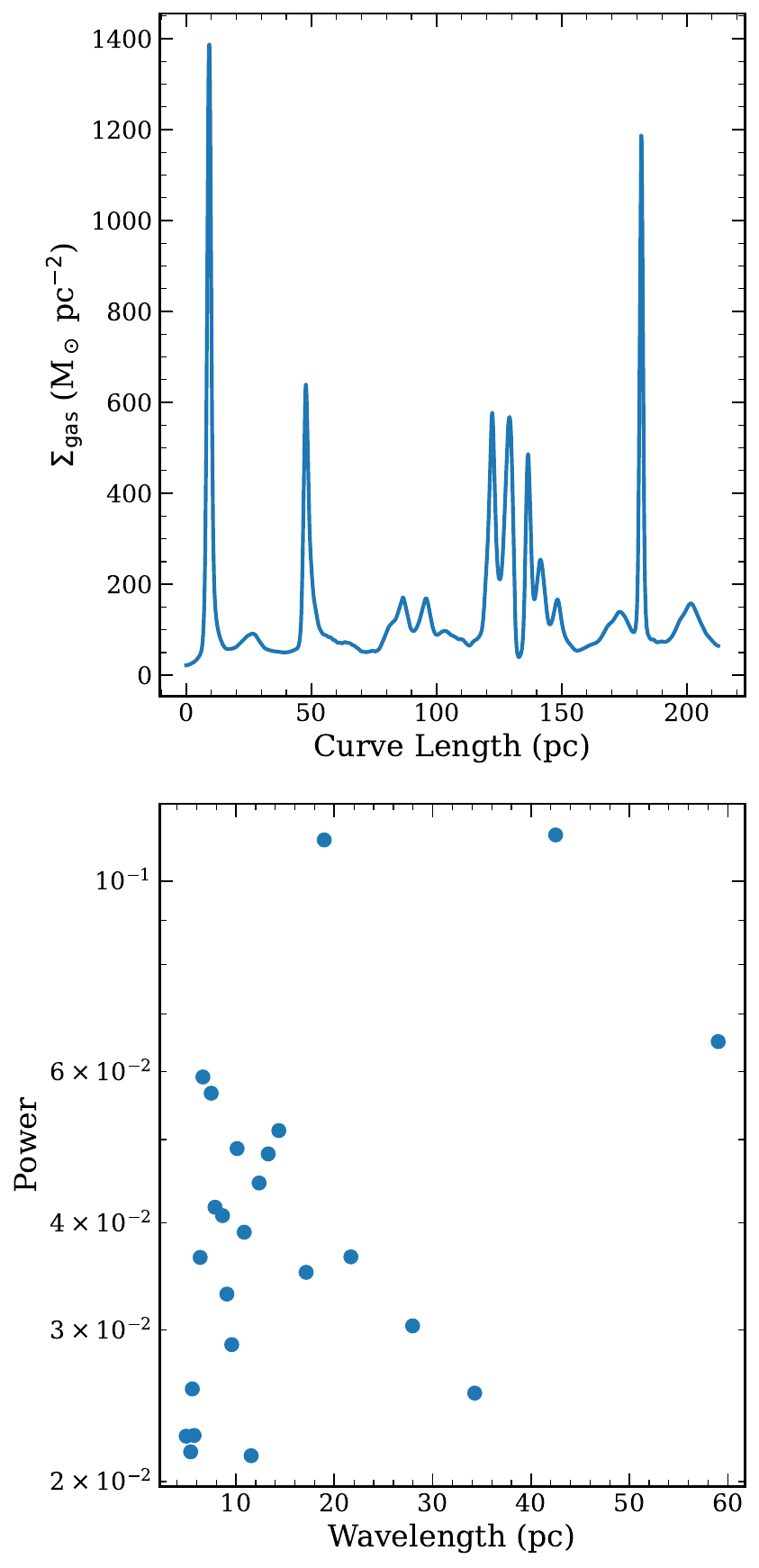}
    \caption{Fragmentation period analysis of the example GMF at t = 730 Myr. Top panel: Smoothed $\Sigma_\mathrm{gas}$ profile along the curve connecting clumps with similar velocity. Bottom panel: Power spectrum from Lomb-Scargle periodogram analysis of the  $\Sigma_\mathrm{gas}$ profile, with statistically significant peaks (confidence level $\geq$ 75\%) marked as blue dots.}
    \label{fig: 12 fragmentation analysis}
\end{figure}

\begin{figure}
    \centering
    \includegraphics[width=\linewidth]{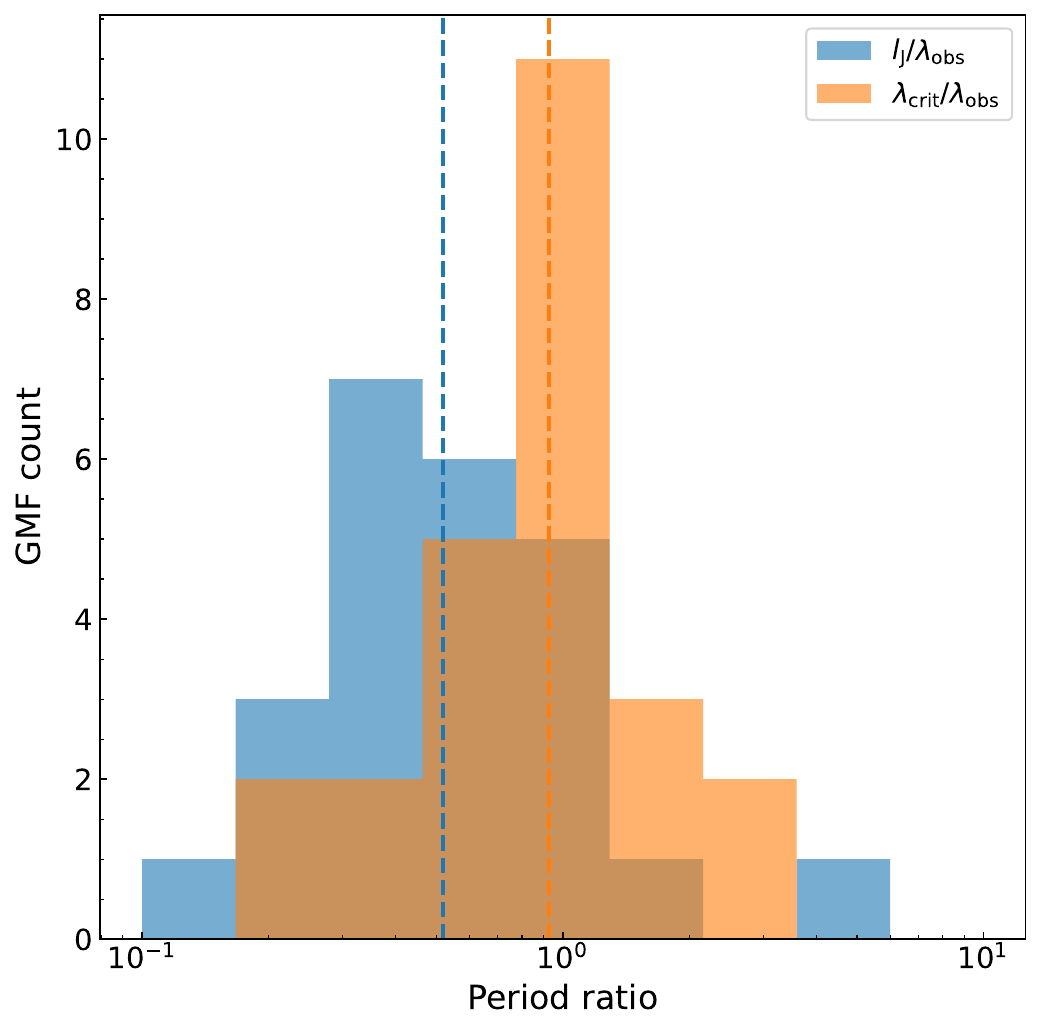}
    \caption{Comparison of theoretical fragmentation periods with observed clump spacing in selected GMFs. The histograms show the distribution of period ratios for two theoretical models: $l_\mathrm{J}/\lambda_\mathrm{obs}$ (blue) and $\lambda_\mathrm{crit}/\lambda_\mathrm{obs}$ (orange). The dashed vertical lines indicate the median values for the distribution with corresponding color.}
    \label{fig: 13 period comparison}
\end{figure}

The smoothed density profile and the fitted oscillation period of the example GMF are shown in \autoref{fig: 12 fragmentation analysis}. From the plot we can see that there exist two main oscillation modes with spatial periods of $\approx 20$ and 40 pc. The reason for multiple modes is that the galactic environment is not exactly the same along the GMF. We take the period with the maximum power as the observed oscillation period $\lambda_\mathrm{obs}$. For each GMF in the sub-sample, we first determine its Jeans length using the mean temperature and mean gas density in the 3D mask surrounding its CoM ($n_\mathrm{H_2,CO} \geq 20 \, \rm cm^{-3}$). We then fit the FWHM of the GMF column density distribution from the $\Sigma_\mathrm{gas}$ map with the \textsc{radfil} package \citep{Zucker2018_RadFil}, and calculate $\lambda_\mathrm{crit}$. We plot the histograms of ratios between theoretical period predictions and $\lambda_\mathrm{obs}$ in \autoref{fig: 13 period comparison}. From the plot we find that $l_\mathrm{J} \sim 0.5 \lambda_\mathrm{obs}$, and $\lambda_\mathrm{crit} \sim 0.9 \lambda_\mathrm{obs}$. Therefore, in the selected GMFs, the sausage instability model almost perfectly predicts the fragmentation period. The fragmentation process of the GMF is controlled by perturbations from MHD turbulence, rather than self-gravity.

Such fragmentation processes create clumps dense enough for local gravitational collapse. By tracing the evolution of the GMFs selected above, we find that for all stars formed inside these GMFs, only 18\% on average form before the snapshot where most clumps are identified, indicating that fragmentation acts as the trigger of star formation inside GMFs. Therefore, magnetic turbulence plays two main roles in GMF evolution: it supports GMFs against global gravitational collapse, while its perturbations enable local star formation.

However, not all GMFs undergo the fragmentation process as in the example shown in \autoref{fig: 11 Fragmentation example}, since cloud–cloud collisions (CCC) are another common scenario in GMF evolution. Such collisions can complicate the morphology of GMFs and interrupt ongoing fragmentation. By tracing the GMCs with $M_\mathrm{CO} > 10^{3} \; \rm M_\odot$ from snapshot $t = 720$ Myr to $t = 730$ Myr, we find that 55.5\% of them merge with other GMCs during this period. For our GMFs, this fraction increases to 72.6\% because of their larger sizes. We conclude that most GMFs experience CCC during their evolution, while the remaining isolated GMFs fragment into dense clumps as predicted by the sausage instability model.

\subsection{GMF destruction}
\label{sec: 4.3 GMF destruction}

For an isolated molecular cloud, stellar feedback from stars formed within the cloud can destroy its molecular mass, and its lifetime can be easily quantified by measuring the ratio of molecular mass over time. However, as our GMFs evolve in a realistic galactic environment, they experience high rates of cloud–cloud collisions (CCC). Such collisions can not only disrupt the filamentary structure of a GMF, but also prolong its existence as a GMC due to mass accretion, making it harder to quantify its lifetime. Here, we provide two time scales to describe the GMF life cycle: the first, $t_\mathrm{fil}$, is the timescale over which a GMF remains filamentary before dispersion, fragmentation, or CCC; the second, $t_\mathrm{H_2}$, is the half-life of gas particles in the identified GMFs remaining molecular.

To determine $t_\mathrm{fil}$, we trace the face-on $\Sigma_\mathrm{gas, CO}$ maps centered on the CoM of each synthetic GMF throughout our simulation snapshots, using map side lengths equal to the GMF length, and determine the total time over which at least one GMF can be identified from these maps using the same criteria as in \autoref{sec:GMF_identify}. From the median of all results, we find $t_\mathrm{fil} = 14^{+2}_{-5}$ Myr, where the uncertainties indicate the 25th and 75th percentiles. With further analysis on the evolution of GMF length and clump numbers, we find that during $t_\mathrm{fil}$, it takes on average $\sim$ 12 Myr for a GMF to grow to its maximum lengths since its formation, then takes $\sim$ 2 Myr to fragment into the largest number of clumps.

\begin{figure}
    \vspace{3mm}
    \centering
    \includegraphics[width=\linewidth]{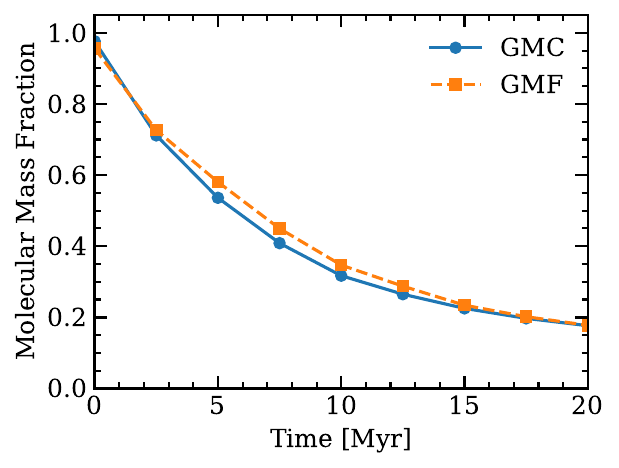}
    \caption{Evolution of molecular mass fraction over time for non-filamentary GMCs (blue) and GMFs (orange).}
    \label{fig: 14 molecular mass ratio}
\end{figure}

As for $t_\mathrm{H_2}$, we use \textsc{despotic} to compute the molecular mass fraction of all gas particles in the GMFs, defining gas particles with at least 50\% molecular mass as being in the molecular phase. We trace these particles from the snapshot at which they are identified to the end of the simulation (20 Myr in total), and determine the molecular mass fraction at each snapshot. A similar analysis is performed for non-filamentary GMCs ($ar < 5$) for comparison, as shown in \autoref{fig: 14 molecular mass ratio}. We find no significant difference in the molecular mass destruction rate between GMCs and GMFs. From the plot, the molecular gas half-life is determined to be $t_\mathrm{H_2} \approx 7$ Myr, roughly half of $t_\mathrm{fil}$. We conclude that GMF destruction occurs on a timescale of 10–15 Myr, driven by turbulent shocks and stellar feedback in our simulation. Our finding that the chemical lifetime of H$_2$ is significantly smaller than the lifetime of the Eulerian molecular structures that those H$_2$ molecules comprise is consistent with the results of \citet{Jeffreson24a}, who observe a similar phenomenon in simulations of a dwarf galaxy.

\section{Discussion}
\label{sec:discussion}

\subsection{GMF identification algorithm}
\label{sec: 5.1 GMF identification algorithm}

As shown in \autoref{fig: 3 length width mass linear mass}, the physical properties of our synthetic GMFs are close to the ``GMF'' catalog in real observations, and we have claimed that the reason for this similarity is primarily that we adopt a column density threshold for GMF identification similar to that used to produce the observed catalog. This raises an obvious question: if we adopt different identification algorithms, can we obtain objects with properties similar to those of other published catalogs? This in turn will help answer the question of whether these catalogs are identifying distinct objects, or whether they are just sampling different parts of the same structures.

\begin{figure*}
\begin{center}
\includegraphics[width=\textwidth]{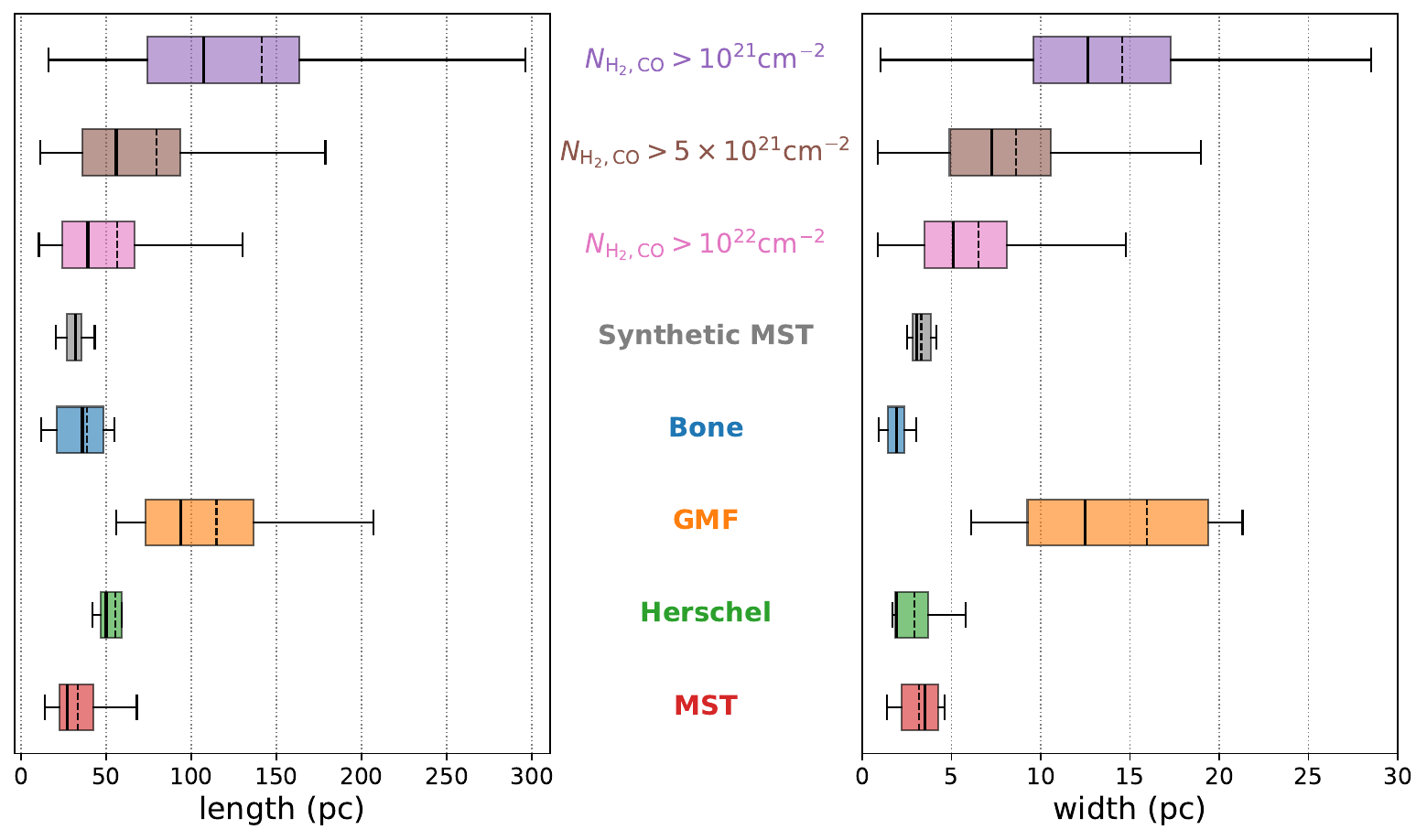}
\includegraphics[width=\textwidth]{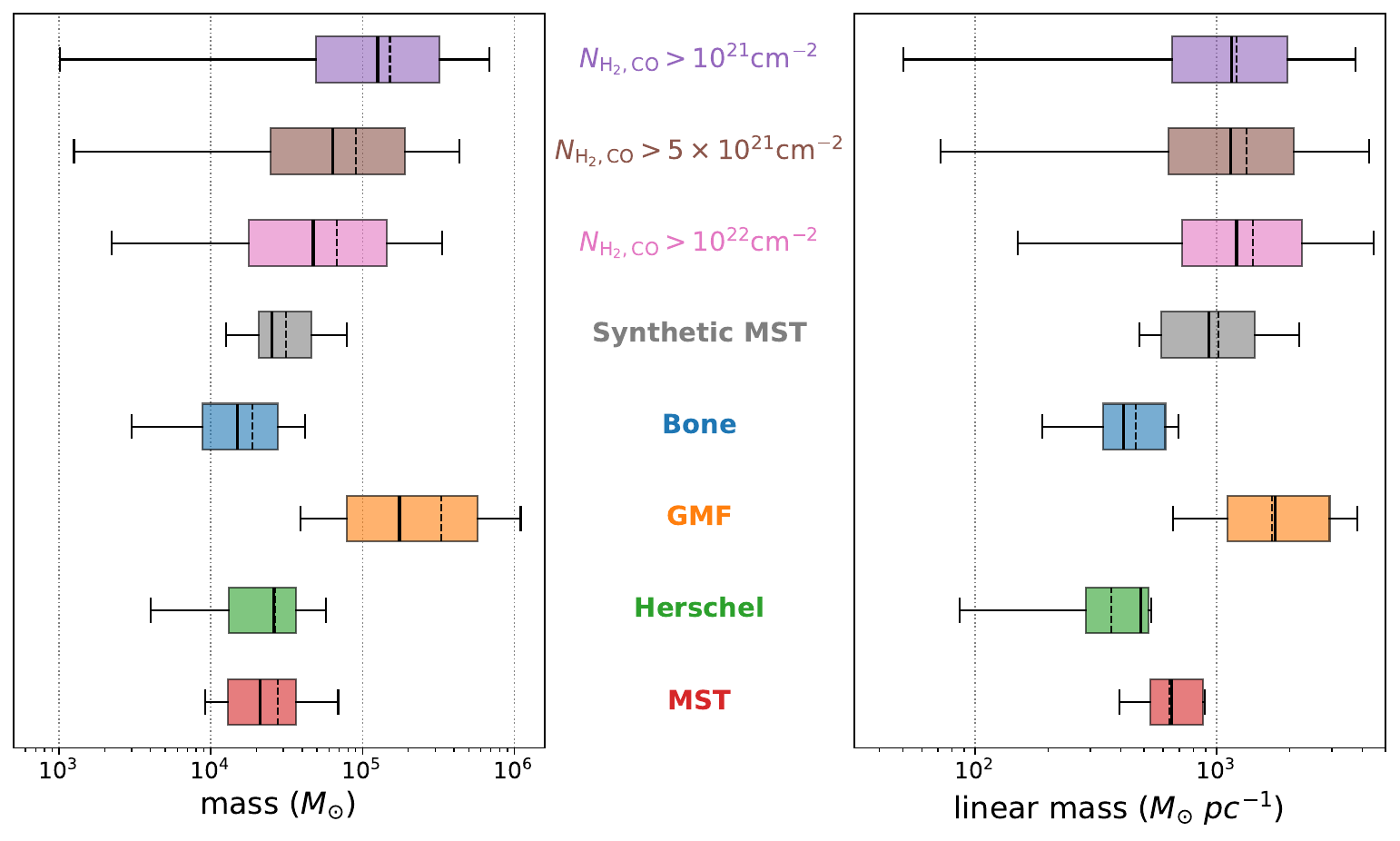}
\caption{Same as \autoref{fig: 3 length width mass linear mass}, but including box-and-whisker plots from synthetic GMFs identified with different $N_\mathrm{H_2}$ thresholds (top three rows) and synthetic GMFs identified with the MST algorithm (fourth row).}
\label{fig: 15 more GMF comparison}
\end{center}
\end{figure*}

To answer these questions, we start with the ``Bone'' and ``Herschel'' catalogs. These objects are identified by visual inspection, and thus lack a universal threshold, but since these structures tend to have smaller sizes and masses compared to those identified as GMFs, we increase the column density threshold in our identification algorithm to $N_\mathrm{H_2, CO} = 5.0 \times 10^{21} \; \rm cm^{-2}$ and $N_\mathrm{H_2, CO} = 1.0 \times 10^{22} \; \rm cm^{-2}$ to test whether structures identified by these thresholds are reasonable matches to observed catalogs. We show the physical properties of the synthetic structures identified with these two higher column density thresholds -- along with the lower threshold of $1.0 \times 10^{22} \; \rm cm^{-2}$ we have adopted up to this point -- as the top three box-and-whisker plots in \autoref{fig: 15 more GMF comparison}. We find that after applying the highest threshold ($N_\mathrm{H_2, CO} = 1.0 \times 10^{22} \; \rm cm^{-2}$), the identified structures match the ``Bone'' and ``Herschel'' catalogs in terms of their lengths, but their widths, masses, and linear masses remain significantly larger. Notably, the linear mass distribution remains nearly unchanged across the three thresholds, indicating that our synthetic structures are self-similar over an order of magnitude in column density. However, this may well be a limitation of our simulation resolution: the ``Bone'' and ``Herschel'' objects are typically $w \sim 1-4$ pc wide and have linear masses of $\mu \sim 300-500$ M$_\odot$ pc$^{-2}$, implying typical densities $\rho \sim \mu/w^2\sim 2-20\times 10^{-21}$ g cm$^{-3}$; for our simulation mass resolution $m_\mathrm{res} = 90$ M$_\odot$, the corresponding linear resolution is $(\rho/m_\mathrm{res})^{-1/3} \sim 0.6 - 1.7$ pc. Thus the widths of objects like those found in the observed ``Bone'' and ``Herschel'' catalogs would be resolved by at most a handful a fluid elements in our simulation, suggesting that our failure to produce objects as thing as the ``Bone'' and ``Herschel'' catalogs may be nothing more than a resolution effect.

The ``MST'' catalog from \cite{Wang2016} is identified using a completely different method: instead of searching for continuous structures in column density maps, filaments are built by connecting adjacent, velocity-consistent dense clumps. To reproduce this catalog from our simulation, we first identify dense clumps from the face-on galactic $\Sigma_\mathrm{gas}$ map following the procedure in \autoref{sec: 4.2 GMF evolution}, and compute the molecular-mass-weighted 3D velocity for each clump. We then build MST trees representing GMFs following the methodology of \cite{Wang2016}:

\begin{enumerate}
    \item Each MST filament contains at least 5 clumps.
    \item The separation between connected clumps is less than 10 pc.
    \item The norm of the velocity difference between connected clumps is less than 2 km s$^{-1}$.
    \item Linearity $f_L > 1.5$, defined as the ratio between the spread (standard deviation) along the MST’s major axis and that along the minor axis.
    \item Total filament length is at least 10 pc.
\end{enumerate}

Using these criteria, we find 17 synthetic MST GMFs from the snapshot at $t = 720$ Myr. Their lengths are measured as the total MST tree length, widths as the mean major-axis lengths of the associated clumps, masses as the total clump masses, and linear masses as mass divided by length. The distributions of these properties are shown as the ``Synthetic MST'' box plots (grey) in \autoref{fig: 15 more GMF comparison}, and generally agree well with the observed MST catalog. The linear masses of the synthetic MSTs show a slightly broader range, which can be narrowed by increasing the clump identification $\Sigma_\mathrm{gas}$ threshold to reduce individual clump mass.

From this comparison, we confirm that our simulation can reproduce the MST catalog from observations. Only 1 of the 17 synthetic MSTs overlaps with the previously identified synthetic GMFs, consistent with the lack of overlap seen between observed catalogs \citep{Zucker2018}. This indicates that MSTs do not trace the high-$\Sigma_\mathrm{gas}$ regions inside continuous structures, but rather the GMFs in their later fragmentation stage. In future work, we plan to trace the evolution of synthetic MSTs backward to build theoretical models capable of recovering the initial conditions of GMFs from the observed clump properties.

\subsection{GMF vs. GMC}
\label{sec: 5.2 GMF vs. GMC}

The galactic simulation results by \cite{Duarte-Cabral2016} suggest that GMFs are an elongated, skinny, and massive subset of GMCs. However, this is largely a consequence of the adopted GMF identification criteria. To study whether GMFs have intrinsic differences from non-filamentary GMCs, we identify GMCs from synthetic face-on CO observations as contours with $N_\mathrm{H_2, CO} \leq 10^{21} \; \rm cm^{-2}$ and aspect ratio $ar < 5$. We then compare three physical properties for both GMFs and GMCs that can be reproduced in observations to test whether these objects are distinct in their density structures, star formation rates, and dynamical timescales: the PDF of $N_\mathrm{H_2, CO}$, the star formation efficiency per free-fall time $\epsilon_\mathrm{ff}$, and the crossing time $t_\mathrm{cross}$.

\begin{figure*}
\begin{center}
\includegraphics[width=\textwidth]{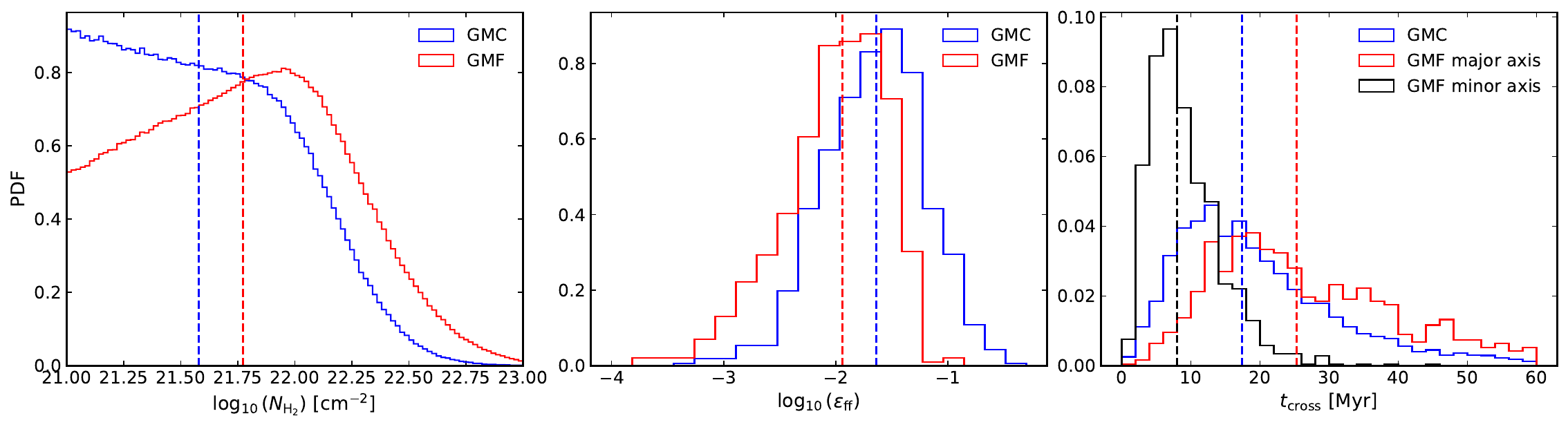}
\caption{PDFs of physical properties for GMCs (blue) and GMFs (red). Left: H$_2$ column density $N_\mathrm{H_2, CO}$. Middle: star formation efficiency per free-fall time $\epsilon_{\rm ff}$. Right: crossing time $t_{\rm cross}$ for GMCs (blue), GMFs along the major axis (red), and GMFs along the minor axis (black). Dashed lines indicate median values for each histogram with corresponding color.}
\label{fig: 17 GMC vs GMF}
\end{center}
\end{figure*}

The $N_\mathrm{H_2, CO}$ PDF is determined from all pixels within the column density masks of all GMCs or GMFs, and plotted in the left panel of \autoref{fig: 17 GMC vs GMF} as blue for GMCs and red for GMFs. The PDF of GMFs is clearly shifted toward higher column densities, likely due to their larger masses or more compact geometries, but the overall shift is relatively small, $\approx 0.25$ dex in the median.

The observed star formation efficiency per free-fall time $\epsilon_\mathrm{ff}$ is calculated following the corrected form in \cite{Hu2021}:
\begin{equation}
    \epsilon_{\rm ff} = \frac{t_{\rm ff}}{t_{\rm dep}} = \sqrt{\frac{3\pi}{32G\rho_g}}\frac{\dot M_*}{M_{\rm gas}},
    \label{eq: star formation efficiency}
\end{equation}
where $t_{\rm ff}$ is the free-fall time, $t_\mathrm{dep}$ the gas depletion time, $\rho_g$ the average gas volume density corrected by the Gini coefficient, $\dot M_*$ the star formation rate, and $M_{\rm gas}$ the total gas mass. For each GMC or GMF contour, we select stellar particles younger than 2 Myr (covering type I and II YSOs) and compute $\dot M_*$ as the total stellar mass divided by 2 Myr. The volume density $\rho_\mathrm{2D}$ is determined from 2D contours assuming spherical GMCs and cylindrical GMFs. The Gini coefficient $g$ of $\Sigma_\mathrm{gas}$ is determined from all pixels within the contour, then used to calculate $\rho_g$ as:
\begin{equation}
    \rho_g = 10^{4.6g - 0.93}\rho_{\rm 2D}.
    \label{eq: rho_g}
\end{equation}
The resulting PDFs of $\epsilon_\mathrm{ff}$ are shown in the middle panel of \autoref{fig: 17 GMC vs GMF}. GMFs have a median $\epsilon_\mathrm{ff}$ that is $\sim 0.3$ dex, or 2 times lower, than GMCs.

This difference likely reflects that in our synthetic observations the objects identified as GMFs are dynamically young and have not yet fully fragmented into dense, star-forming clumps. Because the observed value of $\epsilon_\mathrm{ff}$ represents an integration over past star formation, a selection that picks out predominantly younger objects is likely to yield a systematically lower $\epsilon_\mathrm{ff}$. This is the converse of the more familiar effect whereby selections that pick out evolved regions where much of the gas has been dispersed -- for example catalogs of H~\textsc{ii} regions -- yield artificially high $\epsilon_\mathrm{ff}$ values \citep{Feldmann11a, Krumholz2019}. Both trends are not a reflection of true variation in $\epsilon_\mathrm{ff}$, but rather of the limitations of observational estimators for it that integrate over times long enough that the system can evolve significantly during them.

The crossing time $t_\mathrm{cross}$ is determined for GMCs as the ratio of mean radius to velocity dispersion. For filamentary GMFs, the crossing time is anisotropic. Therefore, for each GMF contour, the crossing time is determined for both the major and minor axes with the corresponding length and velocity dispersion. The resulting PDFs are shown in the right panel of \autoref{fig: 17 GMC vs GMF}. While the distributions differ between GMFs and GMCs, the median $t_\mathrm{cross}$ for GMCs ($\sim 17$ Myr) is comparable to the filamentary lifetime $t_\mathrm{fil}$ of GMFs. This suggests that both GMCs and GMFs evolve morphologically on similar timescales.

Thus the overall picture we form is that GMFs are similar to GMCs, but perhaps represent a slightly earlier and slightly high column density phase of evolution.

\section{Conclusion and future work}
\label{sec:conclusion}

In this work, by conducting a synthetic CO survey of an entire simulated Milky Way–like galaxy, we provide a comprehensive statistical view of the formation and evolution of GMFs. The synthetic GMF catalog successfully reproduces the observed distributions of physical properties from certain observations, demonstrating the realism of our approach and providing a robust framework for comparing theory and observations.

Our analysis shows that GMFs typically evolve through three main stages: formation primarily driven by turbulent compression arising from galactic shear and stellar feedback, fragmentation into dense clumps induced by turbulence perturbations, and eventual dispersal driven by stellar feedback. Cloud–cloud collisions are also frequent during their evolution, significantly reshaping their morphology and often prolonging their lifetimes. These processes highlight that GMFs are not isolated entities, but dynamically evolving structures embedded in a turbulent galactic environment.

We find that magnetized turbulence plays a dual role in GMF evolution. On one hand, it provides support against global gravitational collapse, keeping the GMFs subcritical. On the other hand, its perturbations trigger local fragmentation, leading to clump formation and subsequent star formation. As a consequence, GMFs exhibit systematically lower star formation efficiencies per free-fall time compared to non-filamentary GMCs, reflecting their turbulence-dominated nature.

The typical lifetime of GMFs as filamentary structures is $t_{\mathrm{fil}} = 14^{+2}_{-5}$ Myr, comparable to the GMC crossing time. The molecular gas half-life time, $t_{\mathrm{H2}} \sim 7$ Myr, is similar to that of GMCs, suggesting that while GMFs are transient features, their gas properties evolve on timescales characteristic of the broader molecular cloud population. 


Overall, our results demonstrate that the life cycle of GMFs is dominated by magnetized turbulence, which makes their evolution paths unique from normal GMCs. Due to resolution limits, our simulation cannot reproduce the thinnest GMFs observed in reality. Future high-resolution modeling is needed to comprehensively study the differences between GMFs from different catalogs and identification methodologies. We also plan to integrate MHD simulations across multiple spatial scales to develop theoretical models that can bridge 2D magnetic field observations and 3D magnetic field morphology.

\begin{acknowledgments}
We thank the anonymous referee for their valuable comments on our manuscript.ZH, KW, and KS acknowledge support from the National Key R\&D Program of China and the National Natural Science Foundation of China (NSFC, No. 12573025). ZH acknowledges support from NSFC through grant No. 12503026 and support from Boya Fellowship at Peking University.
KW acknowledges support from China-Chile Joint Research Fund (CCJRF No. 2211) and the Tianchi Talent Program of Xinjiang Uygur Autonomous Region.
MRK acknowledges support from the Australian Research Council through Laureate Fellowship FL220100020. This work was carried out with the assistance of resources from the National Computational Infrastructure (NCI Australia), an NCRIS enabled capability supported by the Australian Government, through award jh2.
\end{acknowledgments}

\begin{contribution}

ZH designed the simulation, performed the data analysis and wrote most parts of this manuscript. KW provided supervision, comparison to observations, and editing. MRK developed simulation data analysis methods and edited the manuscript. KS developed methods to analyze fragmentation periods.


\end{contribution}

%




\appendix

\section{Projection effects on GMF identification}
\label{appendix}

\begin{table}
\centering
\begin{tabular}{lcccccc}
\hline
\hline
& \multicolumn{3}{c}{Edge-on projection} & \multicolumn{3}{c}{Random projection} \\
\cmidrule(lr){2-4} \cmidrule(lr){5-7}
Property & Q1 & Median & Q3 & Q1 & Median & Q3 \\
\hline
Length ratio & 0.99 & 1.04 & 1.27 & 0.91 & 1.04 & 1.38 \\
Width ratio& 0.98 & 1.02 & 1.15 & 0.99 & 1.09 & 1.36 \\
Mass ratio & 0.93 & 1.21 & 1.94 & 0.99 & 1.38 & 2.72 \\
\hline
\end{tabular}
\caption{Percentile statistics for GMF property ratios between the GMF catalog considering foreground and background gas and the original GMF catalog. The table shows the 25th percentile (Q1), median (50th percentile), and 75th percentile (Q3) for length, width, and mass ratios under edge-on (first three columns) and random projection angles (last three columns).
\label{tab:projection_ratios}
}
\end{table}

In \autoref{sec:GMF_identify}, we isolate the GMF by removing foreground and background gas using two perpendicular viewing angles and project it along the minor axis of its face-on contour. However, real observations cannot always easily separate gas structures along the line of sight, and the fixed viewing angle from the Earth can hardly trace the minor axis of each GMF. To assess how these projection effects influence measured GMF properties, we perform two new sets of synthetic observations on our synthetic GMF catalog.

GMFs are typically identified in Position-Position-Velocity (PPV) space in real observations, where gas at different physical depths is blended if it shares a similar $v_\mathrm{LOS}$. To mimic this, we perform synthetic observations using a methodology similar to the in-plane (edge-on) views in \autoref{sec:GMF_identify}. The key difference lies in the gas particle selection: we include all gas particles within 500~pc of the GMF's center of mass, then exclude those outside the $v_\mathrm{LOS}$ range of the GMF identified from the original edge-on view. The 500~pc limit is chosen because gas beyond this distance exhibits a large velocity difference due to galactic rotation. We then identify GMFs from these new synthetic observations using the same criteria in \autoref{sec:GMF_identify}. For each identified GMF, we measure its length, width, and mass, and compute the ratios of these values to the original ones. The 25th, 50th, and 75th percentiles of these ratios are tabulated in the first three columns of \autoref{tab:projection_ratios}. The results show that contaminating LOS gas has a limited effect on measured lengths and widths but introduces a $\sim 20\%$ uncertainty in mass measurements.

We further investigate the effects of random projection angles by repeating the synthetic observations from 100 different in-plane lines of sight, each separated by $3.6\degree$, for each original GMF. We find that $34\%$ of these lines of sight yield no GMF detection under our criteria. For lines of sight with detections, we again compute the property ratios relative to the original catalog; the percentiles are listed in the last three columns of \autoref{tab:projection_ratios}. These values combine the effects of LOS gas and projection angle. The median length and width are affected by less than $10\%$, while the mass can be overestimated by about $40\%$.

We conclude that projection effects would cause approximately one-third of the GMFs in our simulated galaxies to be undetected in real observations. For a detected GMF, its measured length and width are likely close to the intrinsic values, but its mass may be overestimated by $\sim 40\%$. This analysis provides an initial estimate; a more detailed study using an ``Earth-like" vantage point in the simulation to generate high-resolution PPV cubes for GMF and GMC identification is planned for future work.

\bibliography{GMF}{}
\bibliographystyle{aasjournalv7}

\end{document}